\documentclass[conf]{new-aiaa}
\usepackage[utf8]{inputenc}
\usepackage{graphicx}
\usepackage{amsmath}
\usepackage{physics}
\usepackage[version=4]{mhchem}
\usepackage{siunitx}
\usepackage{float}
\usepackage{subcaption}
\usepackage{multirow}
\usepackage{longtable,tabularx}
\title{High-Order Large-Eddy Simulations of a Ducted Wind Turbine}

\author
{Chi Ding
\footnote{Graduate Student in the Department of Mechanical and Aeronautical Engineering, chid@clakrson.edu}
, Bin Zhang
\footnote{Research Assistant Professor in the Department of Mechanical and Aeronautical Engineering, bzhang@clarkson.edu}
, Chunlei Liang
\footnote{Professor in the Department of Mechanical and Aeronautical Engineering, and AIAA Associate Fellow, cliang@clarkson.edu}
, Kenneth D. Visser
\footnote{Associate Professor in the Department of Mechanical and Aeronautical Engineering, AIAA Senior Member, kvisser@clarkson.edu}
, and Guangming Yao
\footnote{Associate Professor in the Department of Mathematics, gyao@clarkson.edu}
}
\affil{Clarkson University, Potsdam, NY, 13676, USA}
\begin{document}

\maketitle

\begin{abstract}
A high-order flux reconstruction method coupled with a high-order sliding mesh method is applied to analyze the performance of a ducted wind turbine at a Reynolds number of $1.25\times 10^6$.  To investigate the impacts of the duct, axial flow simulations are also performed for the corresponding open-rotor turbine. It is shown that the ducted turbine has higher thrust and power outputs than the open rotor. Vorticity and velocity fields of two configurations are also visualized and analyzed to show the differences. To evaluate the effects of yawed flows, simulations are carried out for both turbines under five different yaw angles $\gamma=0^{\circ}, \pm15^{\circ}, \pm 30^{\circ}$. The results reveal that the ducted wind turbine has higher power outputs than the open counterpart at all tested yaw angles.
\end{abstract}

\section{Introduction}
\lettrine{W}{ind} energy has gained growing popularity due to its clean and renewable property. One promising way to harvest wind energy is the ducted wind turbine (DWT), also known as the diffuser-augmented wind turbine (DAWT). Compared with conventional wind turbines, a DWT is equipped with a duct in order to increase power output, and has the potential to exceed the Beltz limit \cite{oka-2013}. To design an efficient DWT, accurate estimates of its loads and the flow fields are of crucial importance. In addition, it is quite often that a DWT operates under yawed inflow conditions when it is surrounded by buildings in urban environment, or is installed with yawed misalignments. There is thus a natural need to study the loads under yawed conditions.

Many efforts have been made to numerically study DWTs. Fletcher \cite{fletcher-1981} used the  Blade Element Momentum (BEM) method to investigate DWTs. Georgalas et al. \cite{georgalas-1991} combined  the lifting-line approach for the rotors and the lifting-surface method for the ducts to study DWTs with large tip clearance. Recently, one of the most popular ways to study DWT is the CFD actuator-disk (CFD-AD) method, which models the rotor by an actuator disk and simulate the flow fields by CFD methods. The first research using the CFD-AD method was done by Hansen et al. \cite{hansen-2000}. This method was also employed to investigate a turbine with a flanged diffuser \cite{ohya-2012}. Venters et al.\cite{venters-2018} utilized the CFD-AD method to study the effects of a duct's geometrical parameters on the rotor's performance. With increasing computational power, it is now possible to perform more accurate CFD simulations of a complete DWT without simplifications of the geometries.

High-order CFD methods (third and above orders) are more accurate and efficient than low-order methods\cite{Z.J.Wang-2013}. The most popular high-order methods include the Discontinuous Galerkin (DG) methods\cite{reed-1973,cockburn-2012}, the Spectral Element (SE) method\cite{patera-1984,karniadakis-2013}, the Spectral Difference (SD) method\cite{kopriva-1996a,kopriva-1996b,kopriva-1998,liu-2006}, and the Flux Reconstruction (FR) method\cite{huynh-2007,huynh-2009}. Among them, the SD and the FR methods are based on the differential form governing equations, and are among the most efficient high-order methods. The FR method has an additional advantage that this scheme provides a unifying framework that many existing high-order schemes, e.g. DG and SD schemes, and even some new schemes can be recovered by choosing different correction functions\cite{huynh-2007}. The FR method has seen a lot of progress recently. Jameson et al.\cite{jameson-2012} performed a nonlinear stability analysis for the FR method. Many researchers have successfully extended FR schemes to mixed elements \cite{castonguay-2011,haga-2010,haga-2011,lu-2012,vincent-2015}. More recent developments about the FR scheme can be found in a review paper\cite{Z.J.wang-2016}.

To deal with flows over rotating objects, Zhang et al. \cite{zhang-2015a, zhang-2015b} introduced the curved dynamic mortar concept and applied it to the development of high-order sliding-mesh SD and FR methods. These methods were later extended to sliding-deforming meshes \cite{zhang-2016a} and to deal with 3D geometries \cite{zhang-2016c}. Zhang et al. \cite{zhang-2018, zhang-2020a, zhang-2020c} further introduced the transfinite mortar concept that has no geometric error and makes a sliding-mesh method arbitrarily high-order accurate in space and high-order in time. This method has been applied to simulate flows around rotating cylinders of different cross-sectional shapes \cite{zhang-2019b}, flapping wings for energy harvesting \cite{zhang-2019a}, and more recently the first high-order eddy-resolving simulation of flow over a marine propeller \cite{zhang-2021}

This study employs the solver developed by Zhang et al. to study the loads and flow fields of a wind turbine in a ducted and an open configurations. The geometries, including the duct and the rotors, are represented accurately using curved high-order meshes. Both axial flow and yawed flow cases are simulated in this work. The rest of paper is organized as follows: Section II introduces the numerical methods including the FR method and the sliding-mesh method based on the concept of dynamic mortar element. Then the simulation setup is given in Section III. Following that, we present and analyze the numerical results in Section IV. Finally, Section V concludes this paper.

\section{Numerical Methods}
\subsection{The Governing Equations}
\subsubsection{The Physical Equations}
We numerically solve the following three-dimensional unsteady Navier-Stokes equations in a conservative form,
\begin{equation}
	\pdv{\mathbf{Q}}{t}  + \pdv{\mathbf{F}}{x} + \pdv{\mathbf{G}}{y} + \pdv{\mathbf{H}}{z} = 0,
	\label{eq:physical}
\end{equation}
where $\mathbf{Q}$ is the vector of conservative variables, $\mathbf{F}$, $\mathbf{G}$, and $\mathbf{H}$ are the flux vectors in each coordinate direction. These terms have the following expressions,
\begin{align}
	\mathbf{Q} & = [\rho ~ \rho u ~ \rho v ~ \rho w ~ E]^{\top},\label{eq:Q}\\[1mm]
	\mathbf{F} & = \mathbf{F}_{\text{inv}} (\mathbf{Q}) +
	\mathbf{F}_{\text{vis}} (\mathbf{Q},\grad{\mathbf{Q}}), \label{eq:F}\\[1mm]
	\mathbf{G} & = \mathbf{G}_{\text{inv}} (\mathbf{Q}) +
	\mathbf{G}_{\text{vis}} (\mathbf{Q},\grad{\mathbf{Q}}), \label{eq:G}\\[1mm]
	\mathbf{H} & = \mathbf{H}_{\text{inv}} (\mathbf{Q}) +
	\mathbf{H}_{\text{vis}} (\mathbf{Q},\grad{\mathbf{Q}}), \label{eq:H}
\end{align}
where $\rho$ is fluid density, $u$, $v$ and $w$ are the velocity components, $E$ is the total energy per volume defined as
$E=p/(\gamma-1)+\frac{1}{2}\rho(u^2+v^2+w^2)$, $p$ is pressure, $\gamma$ is the ratio of specific heats and is set to 1.4. The fluxes have been split into inviscid and viscous parts. The inviscid fluxes are only functions of the conservative variables and have the following expressions,
\begin{equation}
	\mathbf{F}_{\text{inv}} =
	\begin{bmatrix}
		\rho u        \\[1mm]
		\rho u^2 + p  \\[1mm]
		\rho uv       \\[1mm]
		\rho uw       \\[1mm]
		u(E+p)
	\end{bmatrix},\quad
	\mathbf{G}_{\text{inv}} =
	\begin{bmatrix}
		\rho v       \\[1mm]
		\rho uv      \\[1mm]
		\rho v^2 + p \\[1mm]
		\rho vw      \\[1mm]
		v(E+p)
	\end{bmatrix},\quad
	\mathbf{H}_{\text{inv}} =
	\begin{bmatrix}
		\rho w       \\[1mm]
		\rho uw      \\[1mm]
		\rho vw      \\[1mm]
		\rho w^2 + p \\[1mm]
		w(E+p)
	\end{bmatrix}.
	\label{eq:FGHinv}
\end{equation}
The viscous fluxes are functions of the conservative variables and their gradients. The expressions are
\begin{gather}
    \mathbf{F}_{\text{vis}} = -
	\begin{bmatrix}
		0          &
		\tau_{xx}  &
		\tau_{yx}  &
		\tau_{zx}  &
		u\tau_{xx}+ v\tau_{yx} +  w\tau_{zx} + \kappa T_{x}
	\end{bmatrix}^{\intercal},\\
	\mathbf{G}_{\text{vis}} = -
	\begin{bmatrix}
		0          &
		\tau_{xy}  &
		\tau_{yy}  &
		\tau_{zy}  &
		u\tau_{xy} + v\tau_{yy} + w\tau_{zy} +\kappa T_{y}
	\end{bmatrix}^{\intercal},\\
	\mathbf{H}_{\text{inv}} = -
	\begin{bmatrix}
		0           &
		\tau_{xz}   &
		\tau_{yz}   &
		\tau_{zz}   &
		u\tau_{xz} + v\tau_{yz} + w\tau_{zz} +\kappa T_{z}
	\end{bmatrix}^{\intercal},
	\label{eq:FGHvis}
\end{gather}
where $\tau_{ij}$ is shear stress tensor which is related to velocity gradients as $\tau_{ij} = \mu(u_{i,j}+u_{j,i})+ \lambda \delta_{ij}u_{k,k}$, $\mu$ is dynamic
viscosity, $\lambda=-2/3\mu$ based on Stokes' hypothesis, $\delta_{ij}$ is the Kronecker delta, $\kappa$ is thermal conductivity, $T$ is temperature that
is related to density and pressure through the ideal gas law $p=\rho R T$, where $R$ is the gas constant.
\subsubsection{The Computational Equations}
As will be discussed later, we map each moving grid element from the physical space to a stationary standard cubic element in the computational space where the
equations are solved. Assume the mapping is: $t=\tau$, $x=x(\tau,\xi,\eta,\zeta)$, $y=y(\tau,\xi,\eta,\zeta)$ and $z=z(\tau,\xi,\eta,\zeta)$, where
$(\tau,\xi,\eta,\zeta)$ are the computational time and space. It can be shown that the flow equations will take the following conservative form in the computational
space,
\begin{equation}
	\pdv{\widetilde{\mathbf{Q}}}{t} + \pdv{\widetilde{\mathbf{F}}}{\xi} +
	\pdv{\widetilde{\mathbf{G}}}{\eta} + \pdv{\widetilde{\mathbf{H}}}{\zeta} = \mathbf{0},
	\label{eq:computational}
\end{equation}
and the computational variable and fluxes are related to the physical ones through
\begin{equation}
	\begin{bmatrix}
		\widetilde{\mathbf{Q}} \\[1mm]
		\widetilde{\mathbf{F}} \\[1mm]
		\widetilde{\mathbf{G}} \\[1mm]
		\widetilde{\mathbf{H}}
	\end{bmatrix}
	=
	\abs{\mathcal{J}}\mathcal{J}^{-1}
	\begin{bmatrix}
		{\mathbf{Q}} \vphantom{\widetilde{\mathbf{Q}}}\\[1mm]
		{\mathbf{F}} \vphantom{\widetilde{\mathbf{F}}}\\[1mm]
		{\mathbf{G}} \vphantom{\widetilde{\mathbf{G}}}\\[1mm]
		{\mathbf{H}} \vphantom{\widetilde{\mathbf{H}}}
	\end{bmatrix}
	\label{eq:jacobi},
\end{equation}
where $\abs{\mathcal{J}}$ is determinant of the Jacobian matrix and $\mathcal{J}^{-1}$ is the inverse Jacobian matrix, and their expressions are
\begin{equation}
	\abs{\mathcal{J}} =
	\abs{\pdv{(t,x,y,z)}{(\tau,\xi,\eta,\zeta)}} =
	\begin{vmatrix}
		1           & 0          & 0           & 0            \\
		x_{\tau}    & x_{\xi}    & x_{\eta}    & x_{\zeta}    \\
		y_{\tau}    & y_{\xi}    & y_{\eta}    & y_{\zeta}    \\
		z_{\tau}    & z_{\xi}    & z_{\eta}    & z_{\zeta}
	\end{vmatrix},
\end{equation}
\begin{equation}
	\mathcal{J}^{-1} =
	\pdv{(\tau,\xi,\eta,\zeta)}{(t,x,y,z)} =
	\begin{vmatrix}
		1          & 0          & 0           & 0            \\
		\xi_{t}     & \xi_{x}    & \xi_{y}     & \xi_{z}      \\
		\eta_{t}    & \eta_{x}   & \eta_{y}    & \eta_{z}     \\
		\zeta_{t}   & \zeta_{x}  & \zeta_{y}   & \zeta_{z}
	\end{vmatrix}.
\end{equation}
 Besides the flow equations, the Geometric Conservation Law(GCL) \cite{thomas-1979} also needs to be numerically satisfied to ensure free-stream preservation on moving
grids. The GCL equations and the steps for solving them  are described in \cite{zhang-2020a,zhang-2020c,zhang-2021}
\subsection{The Flux Reconstruction Method}
\begin{figure}[H]
	\centering
	\includegraphics[width=2.2in]{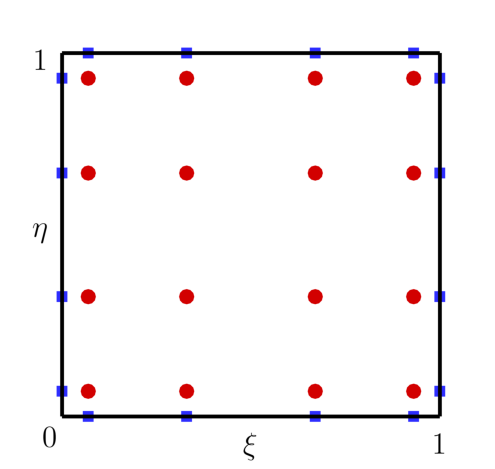}
	\caption{Schematic of solution points (circles) and flux points (squares) in the $\xi-\eta$ plane for a fourth-order FR method.}
	\label{fig:FP_SP}
\end{figure}

The meshes in this work consist of hexahedral elements only. The first step of the FR method is to map each hexahedral element to a standard cubic element of unit size, i.e., $0\leq\xi,\eta,\zeta\leq 1$.This can be done using the following iso-parametric mapping,
\begin{equation}
	\begin{bmatrix}
		x \\[1mm]
		y \\[1mm]
		z
	\end{bmatrix}
	= \sum^{K}_{i=1} M_i(\xi,\eta,\zeta)
	\begin{bmatrix}
		x_i(t)  \\[1mm]
		y_i(t)  \\[1mm]
		z_i(t)
	\end{bmatrix},
\end{equation}
where $K$ is the number of nodes of a hexahedral element, $M_i$ (detailed expressions can be found in \cite{bathe-2006}) is the shape function of the
$i$-th node, and $(x_i,y_i,z_i)$ are the coordinates of the $i$-th node.
Next, solution points (SPs, denoted by $X_s$) and flux points (FPs, denoted by $X_f$) are defined along each coordinate direction in the standard element. Fig. \ref{fig:FP_SP}
shows a schematic of the distribution of the SPs and FPs in the $\xi-\eta$ plane for a fourth-order FR method. Generally, for an $N$-th order FR scheme, there are $N$ SPs and FPs in each direction, where the SPs are in the interior and the FPs are on the boundaries of the standard element. The SPs and FPs are chosen as the Legendre points in this work.
Subsequently, Lagrange interpolation bases are defined at each SP. For instance, at the $i$-th SP we have
\begin{equation}
	h_i(X)= \prod^{N}_{s=1,s\neq i}(\frac{X-X_s}{X_i-X_s}).
\end{equation}
The resulting bases also form a basis of polynomials of degrees less than or equal to $N-1$, i.e., $\boldsymbol{\mathsf{P}}_{N-1}$.These interpolation bases allow the construction
of solution and flux polynomials inside each element through tensor products. For example,
\begin{align}
	\widetilde{\mathbf{Q}}(\xi,\eta,\zeta) = \sum_{i=1}^N \sum_{j=1}^N \sum_{k=1}^N
	\widetilde{\mathbf{Q}}_{ijk} h_i(\xi) h_j(\eta) h_k(\zeta),\label{eq:Qt}\\[1mm]
	\widetilde{\mathbf{F}}(\xi,\eta,\zeta) = \sum_{i=1}^N \sum_{j=1}^N \sum_{k=1}^N
	\widetilde{\mathbf{F}}_{ijk} h_i(\xi) h_j(\eta) h_k(\zeta),
	\label{eq:Ft}
\end{align}
where the subscript $ijk$ denotes the discrete value at the $ijk$-th SP. All these polynomials are in  $\boldsymbol{\mathsf{P}}_{N-1,N-1,N-1}$.
The above solution and flux polynomials are continuous within each element, but discontinuous across cell boundaries. Therefore, common values need to be
defined on cell boundaries. There are various ways to calculate these common values. In this work, the common solution is calculated as the average of the discontinuous values from the two sides of a boundary; the common inviscid fluxes are computed using the Rusanov solver\cite{rusanov-1961}; the common viscous fluxes are computed from the common solutions and common gradients.
There is one more issue: after taking the first-order spatial derivatives in the governing equations, the three flux terms become elements in $\boldsymbol{\mathsf{P}}_{N-2,N-1,N-1}$, $\boldsymbol{\mathsf{P}}_{N-1,N-2,N-1}$, and $\boldsymbol{\mathsf{P}}_{N-1,N-1,N-2}$, respectively, and are inconsistent with the solution term. To fix this issue, the original fluxes need to be reconstructed. This is done by using correction functions that are polynomials of degree no less than $N$. Taking the flux in the $\xi$ direction as an example, the reconstructed flux polynomial is
\begin{equation}
	\widehat{\mathbf{F}} =  \widetilde{\mathbf{F}}(\xi,\eta,\zeta)
	+ [\widetilde{\mathbf{F}}^{\text{com}}(0,\eta,\zeta)
	-  \widetilde{\mathbf{F}}(0,\eta,\zeta)]\cdot g_L(\xi)
	+ [\widetilde{\mathbf{F}}^{\text{com}}(1,\eta,\zeta)
	-  \widetilde{\mathbf{F}}(1,\eta,\zeta)]\cdot g_R(\xi),
\end{equation}
where $\widetilde{\mathbf{F}}$ is form (\ref{eq:Ft}); $\widetilde{\mathbf{F}}^{\text{com}}$ represents the common flux on a cell boundary; $g_L$ and $g_R$ are the left and right correction functions, and are required to satisfy
\begin{equation}
	\begin{alignedat}{2}
		g_{\text{\tiny L}}(0) = 1,\quad g_{\text{\tiny L}}(1) = 0,\\
		g_{\text{\tiny R}}(0) = 0,\quad g_{\text{\tiny R}}(1) = 1,
	\end{alignedat}
\end{equation}
which ensures that
\begin{equation}
	\widehat{\mathbf{F}}(0,\eta,\zeta) = \widetilde{\mathbf{F}}^{\text{com}}(0,\eta,\zeta),\quad
	\widehat{\mathbf{F}}(1,\eta,\zeta) = \widetilde{\mathbf{F}}^{\text{com}}(1,\eta,\zeta)
\end{equation}
i.e., the reconstructed fluxes still take the common values at cell interfaces. In this work, we have employed the $g_{DG}$ correction function\cite{huynh-2007}. The other two fluxes are reconstructed in the same way. Finally, the governing equations can be written in the following residual form,
\begin{equation}
	\eval{\pdv{\widetilde{Q}}{\tau}}_{ijk} = -
	\left[
	\pdv{\widehat{\mathbf{F}}}{\xi}
	+ \pdv{\widehat{\mathbf{G}}}{\eta}
	+ \pdv{\widehat{\mathbf{H}}}{\zeta}
	\right]_{ijk}
	= \mathbf{R}_{ijk}, \quad i,j,k = 1,2,\cdots,N,
\end{equation}
where $\mathbf{R}_{ijk}$ is the residual at the $(i,j,k)$-th SP. This system can be time marched explicitly or implicitly. In the present work, we choose a four-stage three-order explicit Runge-Kutta method. More details can be found in \cite{spiteri-2002,ruuth-2006} .

\subsection{A Sliding Interface Method}
\begin{figure}[H]
	\centering
	\includegraphics[width=2.2in]{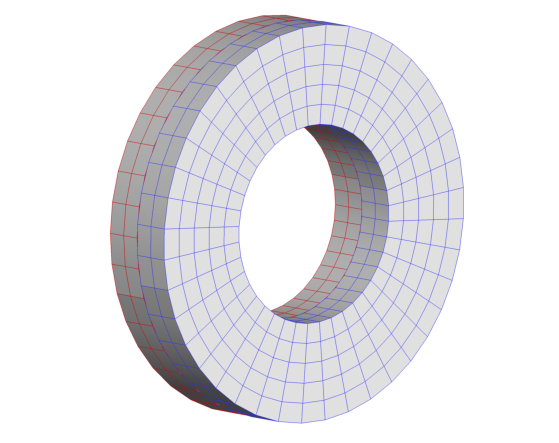}
	\includegraphics[width=2.2in]{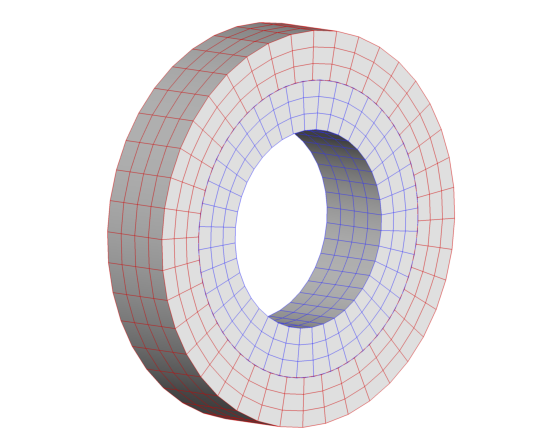}
	\caption{Two types of sliding meshes: left, annular sliding; right, cylindrical sliding.}
	\label{fig:sliding_mesh}
\end{figure}

\begin{figure}[H]
	\centering
	\includegraphics[width=2.2in]{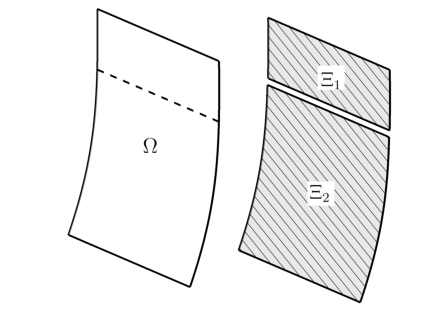}
	\includegraphics[width=0.8in]{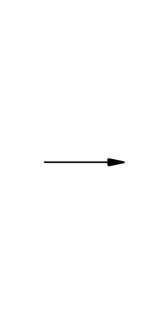}
	\includegraphics[width=2.2in]{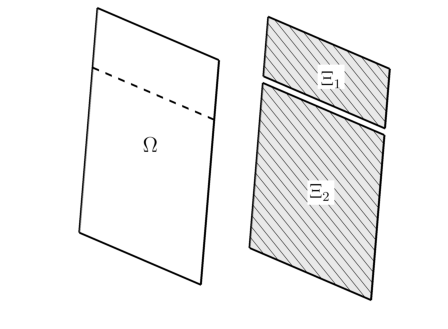}
	\caption{Map curved cell face and mortars to straight ones.}
	\label{fig:mortar_map}
\end{figure}

In three-dimensions, we consider two types of sliding interfaces as depicted in Fig. \ref{fig:sliding_mesh}: one is annular, and the other is cylindrical. To simplify the explanation, we have required the mesh to only unmatch in the azimuthal direction but match in the radial (for annular sliding) or axial (for cylindrical sliding) direction. We also require equal mesh size in the azimuthal direction. These restrictions are imposed for explanation purposes only and can be easily lifted in practice. More details can be found in our previous papers \cite{zhang-2015a,zhang-2015b,zhang-2016a,zhang-2016c,zhang-2018,zhang-2020a,zhang-2020c,zhang-2021}.
Since the SPs on the two sides of a nonconforming sliding interface do not match, we aim to find the best-possible
common values of a variable on the two sides of a sliding interface. This can be achieved by least-squares projections using mortar elements as the intermediate medium. A mortar element is formed by the overlapping region of two cell faces. Taking the cylindrical sliding interfaces as an example, based on the assumptions we have made, a cell face has two mortar elements as sketched in Fig. \ref{fig:mortar_map}. The first step is to map a cell face and the mortars to standard ones as shown in the same figure, using, for example, iso-parametric mapping or transfinite mapping.
Let the $(\xi',\eta')$ denote the mortar space, then the computational and the mortar spaces are related as
\begin{equation}
	\xi = o + s\xi^{\prime},\quad \eta = \eta^{\prime}
\end{equation}
where $0\leq\xi,\eta,\xi,\eta'\leq 1$, and $o$ and $s$ are the offset and scaling of a mortar with respect to a cell face.

\begin{figure}[H]
	\centering
	\begin{subfigure}[b]{0.4\textwidth}
		\includegraphics[width=2.2in]{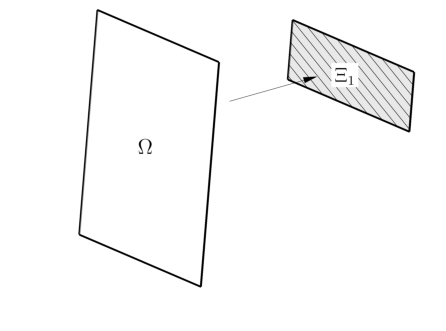}
		\caption{}
		\label{fig:project_1}
	\end{subfigure}
	\begin{subfigure}[b]{0.4\textwidth}
		\includegraphics[width=2.2in]{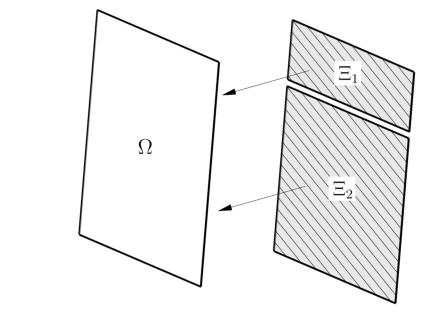}
		\caption{}
		\label{fig:project_2}
	\end{subfigure}
	\caption{Projection between face and mortar: (a) from face to the left side of a mortar, (b) from two mortars back to a face.}
	\label{fig.comparisons}
\end{figure}

Let $\phi$ represent the variable of interest, and obviously it can be represented by the following polynomials on
a cell face $\Omega$ and on the left side of a mortar $\Xi$,
\begin{gather}
	\Phi^{\Omega}(\xi,\eta)   = \sum_{i=1}^N \sum_{j=1}^N \Phi^{\Omega}_{ij} h_i(\xi) h_j(\eta),\\[1mm]
	\Phi^{\Xi,L} (\xi',\eta') = \sum_{i=1}^N \sum_{j=1}^N \Phi^{\Xi,L}_{ij}  h_i(\xi')h_j(\eta'),
\end{gather}
where $\phi^{\Omega}_{ij}$ and $\phi^{\Xi,L}_{ij}$ are the discrete values at the $(i,j)$-th SP on $\Omega$ and the left side of $\Xi$, respectively. The $(\phi^{\Xi,L}_{ij})^,$s are unknown, and can be obtained through the following projection (refer to Fig. \ref{fig:project_1})
\begin{equation}
	\int_0^1 \int_0^1  (\Phi^{\Xi,L}(\xi',\eta') -\Phi^{\Omega}(\xi,\eta))
	h_{\alpha} (\xi') h_{\beta}(\eta') \mathrm{d}\xi' \mathrm{d}\eta' = 0,
	\quad \forall \alpha,\beta = 1,2,...,N.
	\label{eq:project_1}
\end{equation}
Considering the relations in (\ref{eq:project_1}), it can be shown that the above two-dimensional projections is equivalent to the one-dimensional one,
\begin{equation}
	\int_0^1 (\Phi^{\Xi,L}(\xi',X_j) -\Phi^{\Omega}(\xi,X_j))
	h_{\alpha}(\xi')  \mathrm{d}\xi' = 0,
	\quad \forall \alpha = 1,2,...,N.
\end{equation}
where $X_j$ is the coordinate of the $j$-th SP. Evaluating the above equation for all the $\alpha^,$s, we will get a system of equations about $\phi^{\Xi,L}_{1:N,j}$.
Repeating this process for every $j$, we will obtain every $\phi^{\Xi,L}_{ij}$. The values on the right side of a mortar, i.e.,the $(\phi^{\Xi,R}_{ij})^,$, can be obtained in the same way.
After that, we can compute a common value on the mortar, e.g., through averaging or Riemann solver. Let us denote this common values as $\Phi^{\Xi}$. We then
project this common variable back to a cell face from mortars as shown in Fig. \ref{fig:project_2}. And the projection is
\begin{equation}
	\begin{alignedat}{2}
		\int_0^{o^2} &\int_{0}^{1} (\Phi^{\Omega}(\xi,\eta) -\Phi^{\Xi_1}(\xi',\eta'))
		h_{\alpha}(\xi) h_{\beta}(\eta) \mathrm{d}\xi  \mathrm{d}\eta \\[1mm]
		&+ \int_{o^2}^1 \int_{0}^{1} (\Phi^{\Omega}(\xi,\eta) -\Phi^{\Xi_2}(\xi',\eta'))
		h_{\alpha}(\xi) h_{\beta}(\eta) \mathrm{d}\xi  \mathrm{d}\eta = 0
		,\quad \forall \alpha, \beta = 1,2,...,N.
	\end{alignedat}
\end{equation}
where $\phi^{\Omega}$ represents the polynomial of the unknown common variable on face $\Omega$. Similarly, this projection is equivalent
to the following one-dimensional projection,
\begin{equation}
	\int_0^{o^2}  (\Phi^{\Omega}(\xi,X_j)- \Phi^{\Xi_1}(\xi',X_j)) h_{\alpha}(\xi)
	\mathrm{d}\xi
	+ \int_{o^2}^1 (\Phi^{\Omega}(\xi,X_j) - \Phi^{\Xi_2}(\xi',X_j)) h_{\alpha}(\xi)
	\mathrm{d}\xi = 0 ,~ \forall \alpha = 1,2,...,N.
\end{equation}
from which $\phi^{\Omega}_{1:N,j}$ are obtained, and then also every $\Phi^{\Omega}_{ij}$ by representing this process for $j$.

\section{Simulation Setup}
\subsection{Geometry}
\begin{figure}[h!]
	\centering
	\includegraphics[width=1.8in]{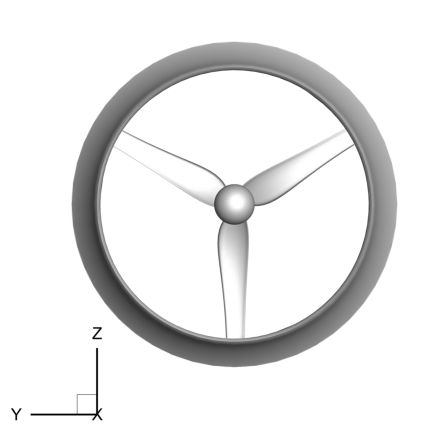} \qquad
	\includegraphics[width=1.8in]{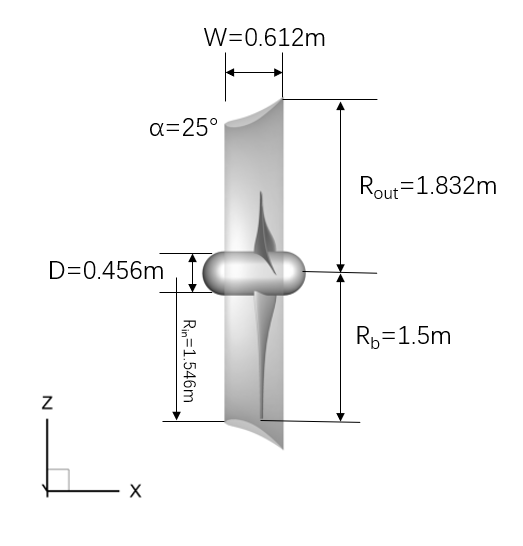}
	\caption{The ducted wind turbine geometry: left, front view; right, lateral view.}
	\label{fig:dwt_geo}
\end{figure}

The ducted wind turbine was designed by Prof. Visser at Clarkson University. Fig. \ref{fig:dwt_geo} illustrates the duct and rotor in both the front and lateral views. As shown in Fig. \ref{fig:dwt_geo}, the inlet radius of the duct is 0.456m, and the exit radius is 1.832m. For the duct section, the width is 0.612m, and the angle of attack is $25 ^\circ$.  The rotor has three 1.5m long blades, and the cylindrical hub is closed by two hemispherical ends of which the diameter is 0.456m. The rotor in the open configuration has the same size as that of the ducted configuration.
\subsection{Computational Mesh}
\begin{figure}[h!]
	\centering
	\begin{subfigure}[b]{0.4\textwidth}
		\centering
		\includegraphics[width=1.8in]{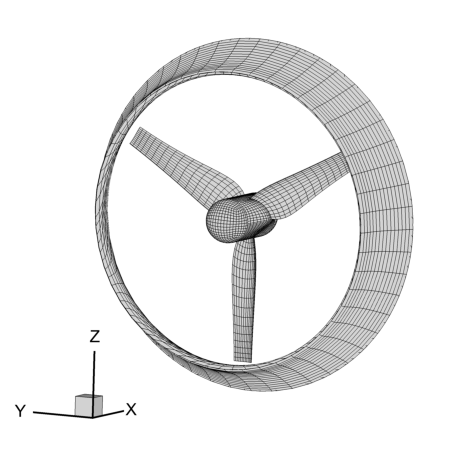}
		\caption{}
		\label{fig:dwt_surface}
	\end{subfigure}
	\begin{subfigure}[b]{0.4\textwidth}
		\centering
		\includegraphics[width=1.8in]{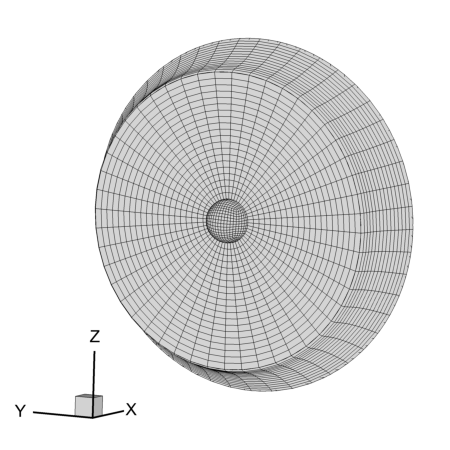}
		\caption{}
		\label{fig:dwt_sliding}
	\end{subfigure}
	\caption{Computational mesh of the ducted configuration: (a) rotor and duct, (b) sliding interface.}
	\label{fig:dwt_surface_mesh}
\end{figure}

\begin{figure}[h!]
	\centering
	\begin{subfigure}[b]{0.4\textwidth}
		\centering
		\includegraphics[width=1.8in]{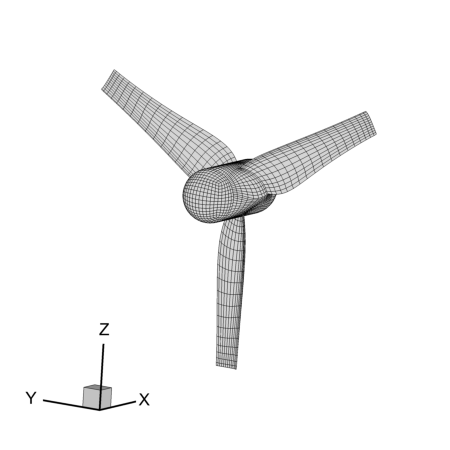}
		\caption{}
		\label{fig:owt_surface}
	\end{subfigure}
	\begin{subfigure}[b]{0.4\textwidth}
		\centering
		\includegraphics[width=1.8in]{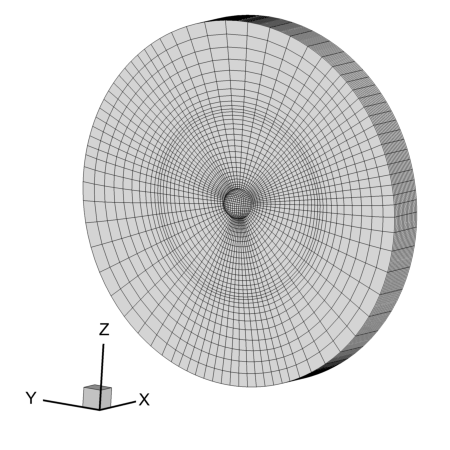}
		\caption{}
		\label{fig:owt_sliding}
	\end{subfigure}
	\caption{Computational mesh of the open-rotor configuration: (a) rotor, (b) sliding interfaces.}
	\label{fig:owt_surface_mesh}
\end{figure}

\begin{figure}[h!]
	\centering
	\begin{subfigure}[b]{0.32\textwidth}
		\includegraphics[width=1.8in]{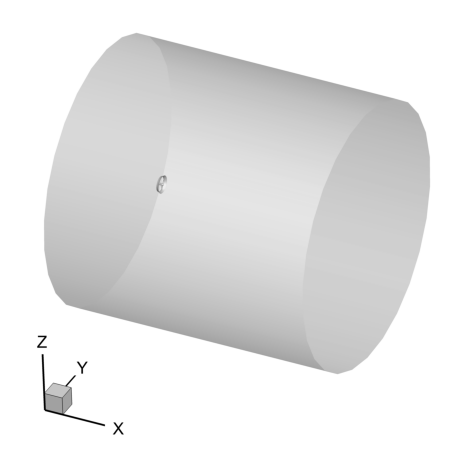}
		\caption{}
		\label{fig:overall}
	\end{subfigure}
	\begin{subfigure}[b]{0.32\textwidth}
		\includegraphics[width=1.8in]{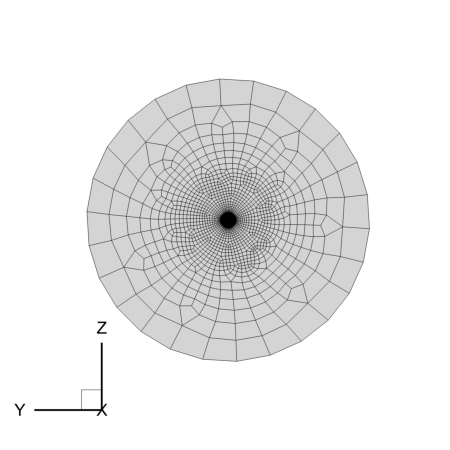}
		\caption{}
		\label{fig:front}
	\end{subfigure}
	\begin{subfigure}[b]{0.32\textwidth}
		\includegraphics[width=1.8in]{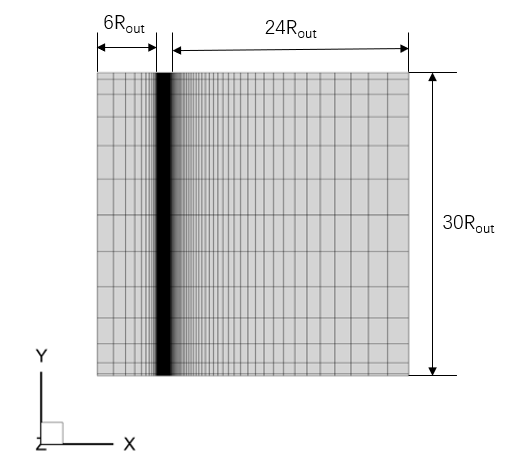}
		\caption{}
		\label{fig:side}
	\end{subfigure}
	\caption{Computational mesh of the ducted wind turbine: (a) overall view, (b) front view (c) lateral view.}
	\label{fig:dwt_mesh}
\end{figure}

In this study, unstructured hexahedral meshes are used for the simulation. Each computation mesh consists of a rotating part and a stationary part. The rotating part includes the rotor. Fig. \ref{fig:dwt_surface_mesh} shows the surface mesh of the duct, the rotor, and the front sliding interface, for the ducted case. For this case, the rotating subdomain has two sliding interfaces. One interface is visible in Fig. \ref{fig:dwt_sliding}, and the other interface is behind the duct. The height of the first mesh layer on the blades is set to $6\times10^{-3}R_{out}$, and the first off-wall solution point is located $2.8\times10^{-4}R_{out}$ away from the blades (for the fifth-order scheme).
Fig. \ref{fig:owt_surface_mesh} illustrates the surface meshes on the rotor and sliding interfaces for the open-rotor configuration. It can be seen that the mesh distribution of the rotor is almost the same as that of the ducted one, and this rotating subdomain has three sliding interfaces: two in the axial direction, and one in the azimuthal direction. The radius of this rotating part is $1.5R_{out}$, and is larger than that of the ducted case because we need more space to achieve a smooth transition from the blade tip to the azimuthal sliding interface on which an orthogonal mesh distribution is needed.
Both the ducted and open-rotor cases have the same cylindrical farfield as shown in Fig. \ref{fig:dwt_mesh}. The inlet is located $6 R_{out}$ ahead of the rotor, and its radius is $30R_{out}$.  The outlet is placed $24R_{out}$ behind the rotor. The blockage ratio is $0.44\%$. For the ducted case, the rotating part has 29,469 hexahedral elements, and the stationary part has 262,941 elements. In total, the mesh has 292,410 elements. A fifth-order FR scheme in space is employed, and the resulting number of degrees of freedom is $36.6$ million. The open case has 38,454 hexahedral elements on the rotating subdomain, and 258,621 elements on the stationary subdomain. The resulting number of degrees of freedom is $37.1$ million.

For the boundary conditions, the inlet is treated as a Dirichlet boundary, the outlet, and the outer boundary as characteristic farfields. The duct and blade surfaces are set as no-slip adiabatic walls. The freestream Mach number $Ma_{\infty}$ is set to 0.08 to ensure that the compressibility effect is negligible.

\subsection{Nondimensional Parameters}
\begin{table}[h!]
	\caption{Operating conditions.}
	\centering
	\begin{tabular}{cc}
		\hline                               			\\[-3ex]
		Parameter            &  Value					\\
		\hline											\\[-2ex]
		Freestream velocity      &  10 m/s       			\\[1ex]
		Kinematic vicosity   &  $1.4616\times10^{-5}$  	\\[1ex]
		Rotation axis        &  x-direction            	\\[1ex]
		Rotation speed       &  -26.18 rad/s  			\\
		\hline
	\end{tabular}
	\label{tab.para}
\end{table}
The operating conditions for the wind turbine are listed in Table \ref{tab.para}. Based on this table, two governing nondimensional parameter can be obtained. One is the Reynolds' number $Re = U_{\infty}R/v = 1.25\times10^6$, where $U_{\infty}$ is the freestream velocity, $R$ is the duct's exit radius, and $v$ is kinematic viscosity.
Since the Reynolds number is very high, the flow around the turbine will become quite turbulent, and the simulation may become unstable due to the aliasing error\cite{jameson-2012}, To reduce the instability, a filter reported in \cite{fischer-2001} (with  strength $\alpha=0.5$) is employed during the simulation.

Another governing parameter is the nondimensional angular speed $\omega^*=\omega R/U_{\infty} = -4.79$. For the present work, the nondimensional time step size is  $\Delta t R/U_{\infty} = 2.5\times 10^{-5}$, which means that the turbine rotates 0.0069 degrees per time step and thus the turbine's rotation can be simulated accurately.

All the simulations are run over $37.5$ time units (about 27 revolutions). From $t R/U_{\infty} = 25$, we start to collect data to compute the averaged flow field. The averaging calculation is performed over around $9.5$ revolutions. For the rotating part, we can obtain the phase-averaged statistics. For the stationary part, the time-averaged statistics can be computed.

\section{Results and Discussions}
\subsection{Axial Flow}
\subsubsection{Load Analysis}
To study the loads on the turbine quantitatively, two performance characteristics are calculated: thrust coefficient $C_T$, and power coefficient $C_P$. They are defined as below.

\begin{gather}
	C_T = \frac{T}{\frac{1}{2}\rho U_{\infty}^2 A_{rot}} \\[1ex]
	C_P = \frac{Q\omega}{\frac{1}{2}\rho U_{\infty}^3 A_{rot}}
\end{gather}
where $T$ is the thrust acted on the rotor, $Q$ is the torque in the x-direction, and $A_{rot}$ is the area of the rotor rotation plane. It should be noted that in this study we only consider the thrust acted on the rotor rather than on all the components.
\begin{figure}[h!]
	\centering
	\begin{subfigure}[b]{0.67\textwidth}
		\includegraphics[width=4.5in]{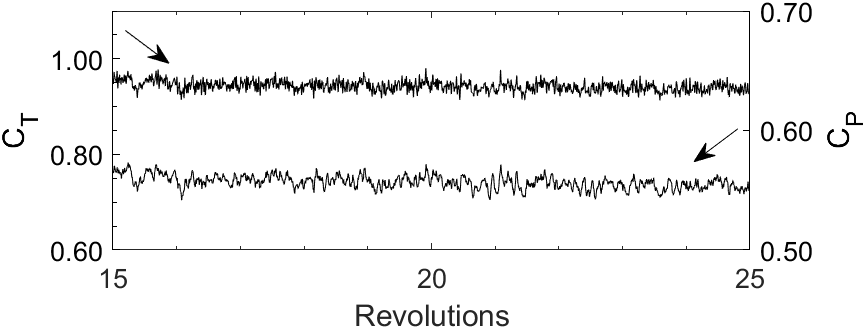}
		\caption{Ducted}
		\label{fig:dwt_perform}
	\end{subfigure}\\[2ex]
   	\begin{subfigure}[b]{0.67\textwidth}
   		\includegraphics[width=4.5in]{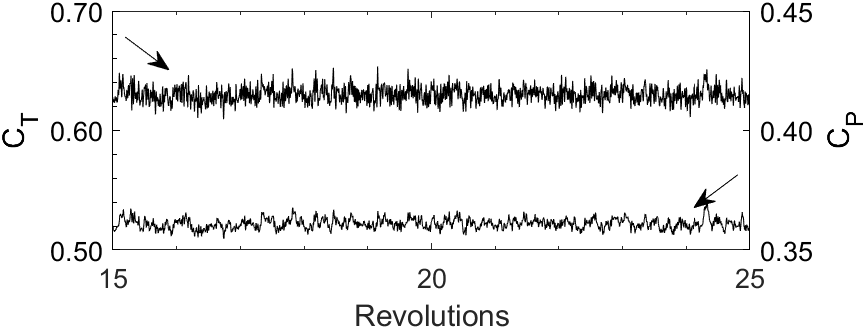}
   		\caption{Open-rotor}
   		\label{fig:owt_perform}
   	\end{subfigure}
   	\caption{Time histories of loads in the two configurations}
   	\label{fig:perform}
\end{figure}
Fig. \ref{fig:perform} demonstrates the time histories of $C_T$ and $C_P$ for the two configurations. In both cases, thrust and power outputs have reached a steady state  with high-frequency fluctuations caused by the turbulent flows.

Table \ref{tab.load} lists the mean values and the r.m.s (root-mean-square) deviations of $C_T$ and $C_P$, their pressure and viscous components. For both configurations, the pressure contributions for $C_T$ and $C_P$ are dominant, and the r.m.s values are about two magnitudes smaller than corresponding mean values. It can be seen that  the rotor in the ducted case has larger mean loads than its open-rotor counterpart. Both the mean $C_T$ and the mean $C_P$ of the ducted case are  about $50\%$ higher than that of the open-rotor case. The ducted case also has a higher root-mean-square deviations for $C_T$ and $C_P$ than the open-rotor case, i.e., the load fluctuations for the ducted one are stronger.

\begin{figure}[h!]
	\centering
	\begin{subfigure}[b]{0.35\textwidth}
		\centering
		\includegraphics[width=2.0in]{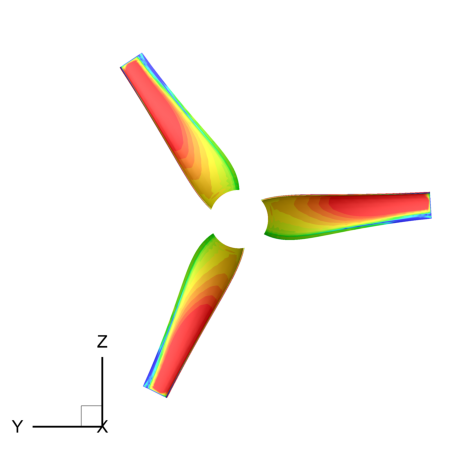}
		\caption{upstream}
		\label{fig:dwt_p_up}
	\end{subfigure}
	\begin{subfigure}[b]{0.35\textwidth}
		\centering
		\includegraphics[width=2.0in]{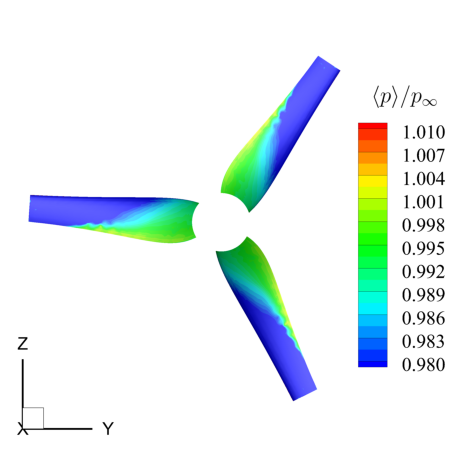}
		\caption{downstream}
		\label{fig:dwt_p_down}
	\end{subfigure}
	\caption{Phase-averaged pressure contours of the blades in the ducted configuration.}
	\label{fig:dwt_p}
\end{figure}

\begin{figure}[h!]
	\centering
	\begin{subfigure}[b]{0.35\textwidth}
		\centering
		\includegraphics[width=2.0in]{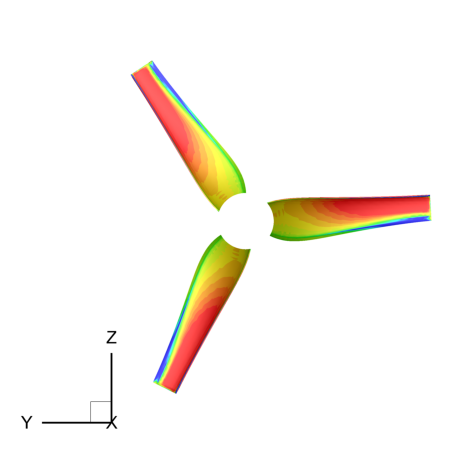}
		\caption{upstream}
		\label{fig:owt_p_up}
	\end{subfigure}
	\begin{subfigure}[b]{0.35\textwidth}
		\centering
		\includegraphics[width=2.0in]{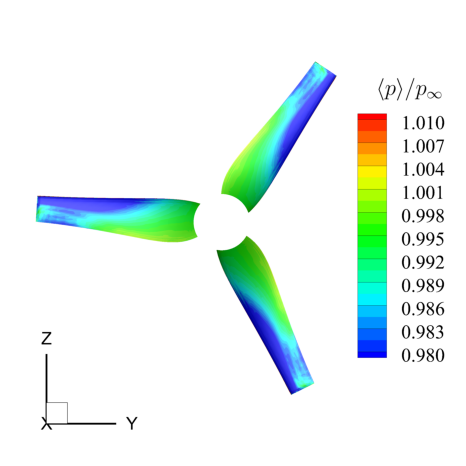}
		\caption{downstream}
		\label{fig:owt_p_down}
	\end{subfigure}
	\caption{Phase-averaged pressure contours of the blades in the open-rotor configuration.}
	\label{fig:owt_p}
\end{figure}

\begin{table}[h!]
	\caption{Loads for the ducted and open-rotor cases.}
	\centering
	\begin{tabular}{cccccccc}
		\hline
		&  & $C_T$ & $C_{T,p}$ & $C_{T,v}$ &  $C_{P}$ & $C_{P,p}$ & $C_{P,v}$\\
		\hline \\[-2ex]
		\multirow{ 2}{*}{Ducted} & mean  & 9.47E-01  & 9.47E-01  & 1.97E-04 & 5.60E-01 & 5.61E-01 & -1.19E-03\\[1ex]
		                         & r.m.s & 1.43E-02  & 1.43E-02  & 1.46E-06 & 8.30E-03 & 8.29E-03 & 1.25E-05 \\[1ex]
		\hline \\[-2ex]
		\multirow{ 2}{*}{Open-rotor} & mean  & 6.30E-01  & 6.30E-01  & 1.97E-04  & 3.61E-01  & 3.62E-01 & -1.31E-03 \\[1ex]
		                             & r.m.s & 6.13E-03  & 6.13E-03  & 8.28E-07  & 2.03E-03  & 2.03E-03 &  4.81E-06 \\[1ex]
		\hline
	\end{tabular}
	\label{tab.load}
\end{table}
To better understand the load distributions on the blades, we plot the phase-averaged pressure contours on the blade surfaces in Figs. \ref{fig:dwt_p} and \ref{fig:owt_p}. For the ducted configuration, the upstream face experiences a higher pressure than the downstream face. Thus the rotor experience a drag force pointed towards x-direction. The outboard part of blades, especially the area around the leading edge, has a higher pressure difference, which means the loads concentrate in this region.

\begin{figure}[h!]
	\centering
	\begin{subfigure}[b]{0.3\textwidth}
		\centering
		\includegraphics[width=2.0in]{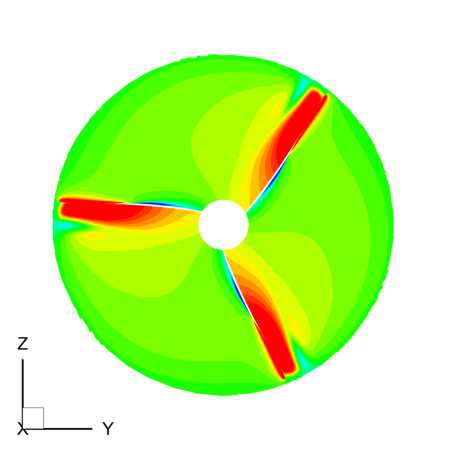}
		\caption{x=0.03R}
		\label{fig:dwt_p_r_0.02}
	\end{subfigure}
	\begin{subfigure}[b]{0.3\textwidth}
		\centering
		\includegraphics[width=2.0in]{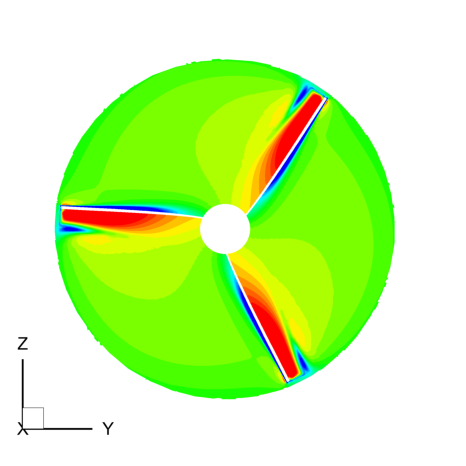}
		\caption{x=0.04R}
		\label{fig:dwt_p_r_0.04}
	\end{subfigure}
	\begin{subfigure}[b]{0.3\textwidth}
		\centering
		\includegraphics[width=2.2in]{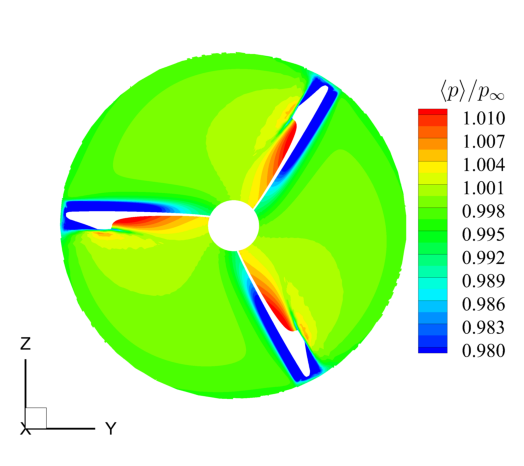}
		\caption{x=0.05R}
		\label{fig:dwt_p_r_0.05}
	\end{subfigure}
	\caption{Phase-averaged pressure contours of the ducted configuration at different streamwise locations.}
	\label{fig:dwt_p_r}
\end{figure}

\begin{figure}[h!]
	\centering
	\begin{subfigure}[b]{0.3\textwidth}
		\centering
		\includegraphics[width=2.0in]{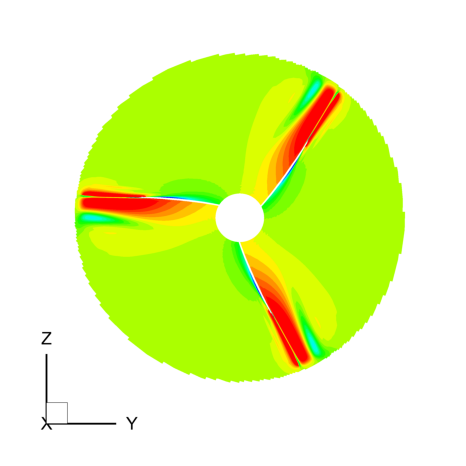}
		\caption{x=0.03R}
		\label{fig:owt_p_r_0.03}
	\end{subfigure}
	\begin{subfigure}[b]{0.3\textwidth}
		\centering
		\includegraphics[width=2.0in]{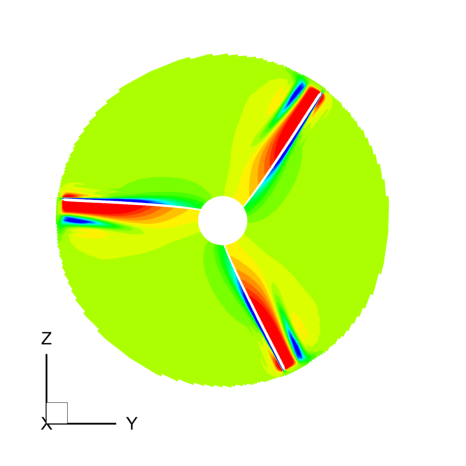}
		\caption{x=0.04R}
		\label{fig:owt_p_r_0.04}
	\end{subfigure}
	\begin{subfigure}[b]{0.3\textwidth}
		\centering
		\includegraphics[width=2.2in]{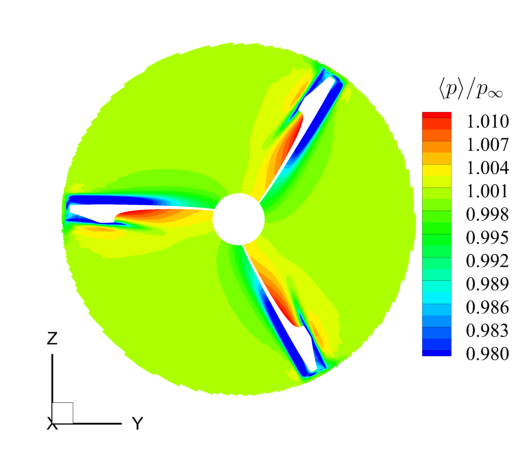}
		\caption{x=0.05R}
		\label{fig:owt_p_r_0.05}
	\end{subfigure}
	\caption{Phase-averaged pressure contours of the open-rotor configuration at different streamwise locations.}
	\label{fig:owt_p_r}
\end{figure}

For comparison, the open rotor's phase-averaged pressure contours are given in Fig. \ref{fig:owt_p}. A similar pressure distribution as the ducted one is observed. One major difference is that the downstream face has smaller low pressure area than that of the ducted configuration, which leads to a lower $C_T$ than the ducted one.
For the ducted case, the phase-averaged pressure contours at different streamwise locations are shown in Fig. \ref{fig:dwt_p_r}. It can be seen that due to the pressure difference between two sides of the blade, all three blade sections experience a torque pointed towards the negative x-axis. This torque direction is the same as the rotation direction, which means a positive work is done on the rotor and the wind energy is transferred to the rotor's mechanical energy. High pressure difference appears on the outboard region, which means most of torque comes from this region.
We also plot the phase-averaged pressure field around the open rotor in Fig. \ref{fig:owt_p_r}. A similar trend for pressure distribution is observed. The rotor
in this configuration also experiences a torque in the same direction as its rotation, and has the main load on the blades' outboard part.
\subsubsection{Wake Dynamics}
\begin{figure}[H]
	\centering
	\includegraphics[width=4.5in]{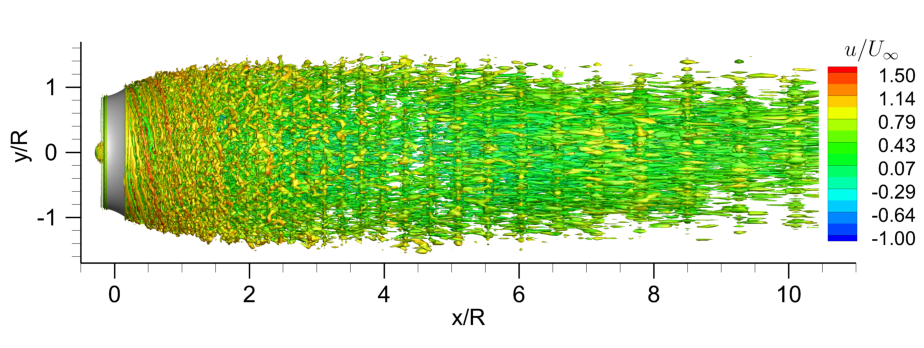}	\\
	\includegraphics[width=4.5in]{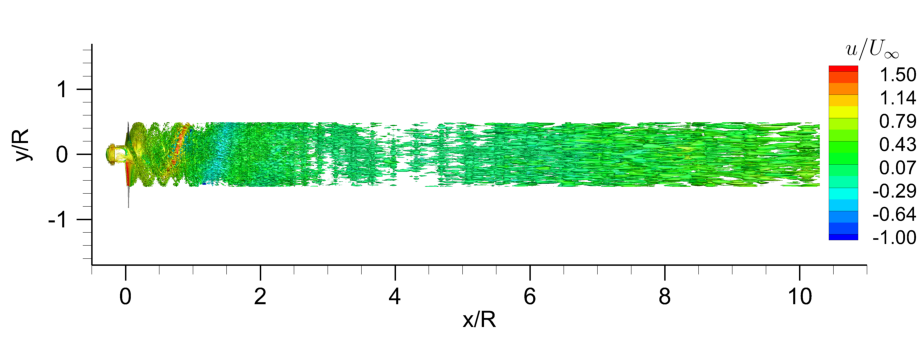}
	\caption{Isosurface of instantaneous Q-criterion $Q_{cr}R^2/U_{\infty}=15$ of the ducted case.}
	\label{fig:dwt_Q_ins}
\end{figure}

\begin{figure}[H]
	\centering
	\includegraphics[width=4.5in]{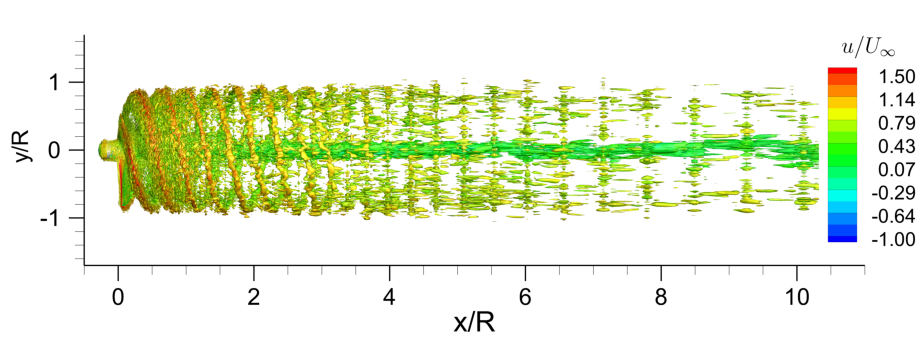}	\\
	\includegraphics[width=4.5in]{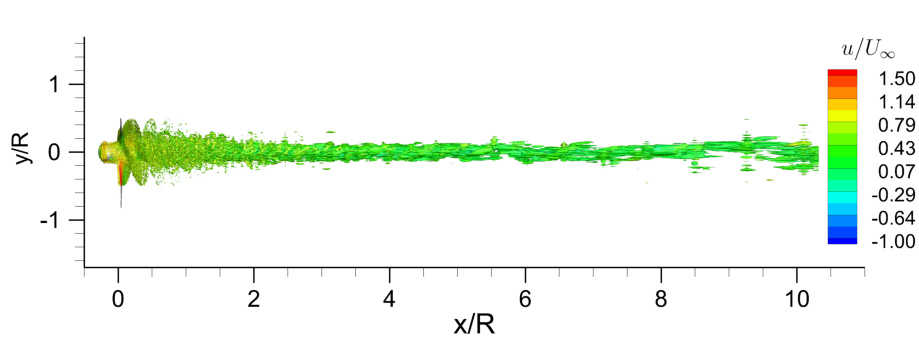}
	\caption{Isosurface of instantaneous Q-criterion $Q_{cr}R^2/U_{\infty}=15$ of the open-rotor case.}
	\label{fig:owt_Q_ins}
\end{figure}

To investigate the wake dynamics of the two configurations, isosurfaces of instantaneous Q-criterion colored by the streamwise velocity are plotted in Figs. \ref{fig:dwt_Q_ins} and \ref{fig:owt_Q_ins}. For the ducted case, as shown in Fig.\ref{fig:dwt_Q_ins}, a lot of vortices are shed from the duct's inner surface. The wake area is wrapped by these small vortex structures from the duct exit to $x=10R$. To obtain more details, in Fig. \ref{fig:dwt_Q_ins}, we also plot isosurfaces of Q-criterion with the area $r>=0.5R$ blanked. It is seen that a lot of trailing vortices are generated, and they are well resolved even up to $x=10R$, which means the wake area is filled with small flow structures.
Fig.\ref{fig:owt_Q_ins} presents Q-criterion isosurfaces of the open-rotor case. As can be seen, this case has much stronger and more coherent tip vortices compared with the ducted case. When the area $r>=0.5R$ is blanked, we can clearly see that a hub vortex is formed just behind the hub end and remains stable for a very long distance. Compared with the ducted case, the flow around the open rotor does not have strong vortices shedding off from blade trailing edges, and thus has a load with smaller r.m.s deviation values.

\begin{figure}[h!]
	\centering
	\begin{subfigure}[b]{0.48\textwidth}
		\centering
		\includegraphics[width=3.0in]{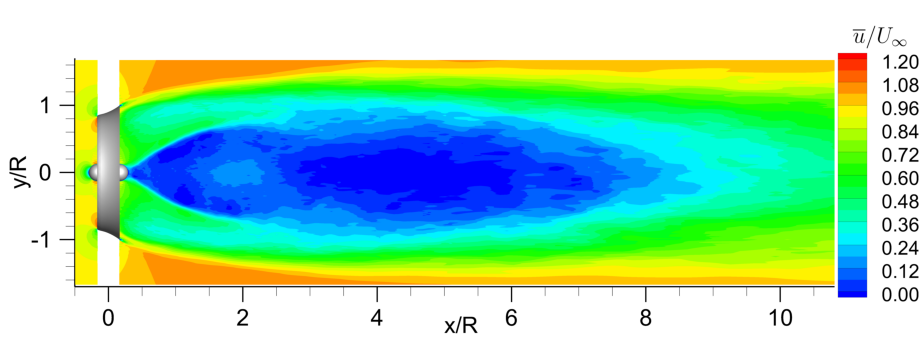}
		\caption{Dcuted}
		\label{fig:dwt_ave_u_stream}
	\end{subfigure}
	\begin{subfigure}[b]{0.48\textwidth}
		\centering
		\includegraphics[width=3.0in]{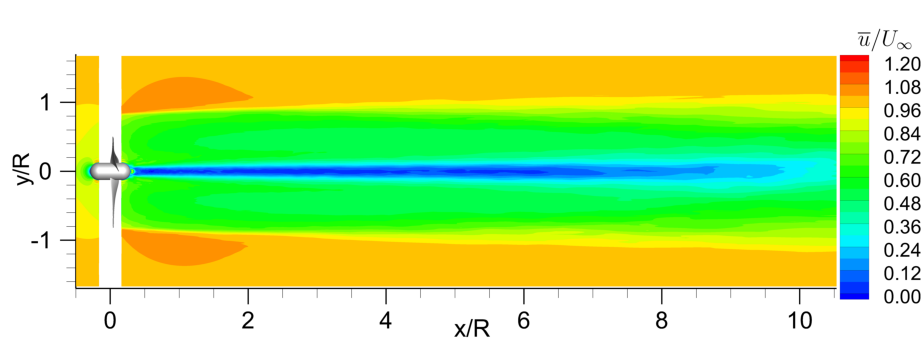}
		\caption{Open-rotor}
		\label{fig:owt_ave_u_stream}
	\end{subfigure}
	\caption{Time-averaged streamwise velocity contours in the central x-y plane}
	\label{fig:ave_u_stream}
\end{figure}

\begin{figure}[h!]
	\centering
	\begin{subfigure}[b]{0.48\textwidth}
		\centering
		\includegraphics[width=3.0in]{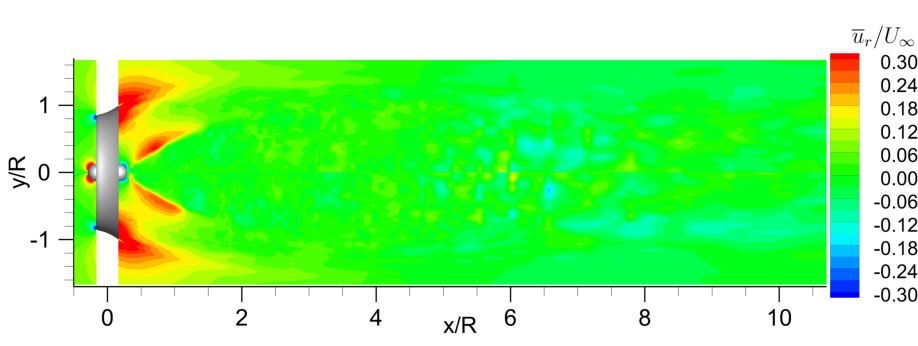}
		\caption{Dcuted}
		\label{fig:dwt_ave_u_radial}
	\end{subfigure}
	\begin{subfigure}[b]{0.48\textwidth}
		\centering
		\includegraphics[width=3.0in]{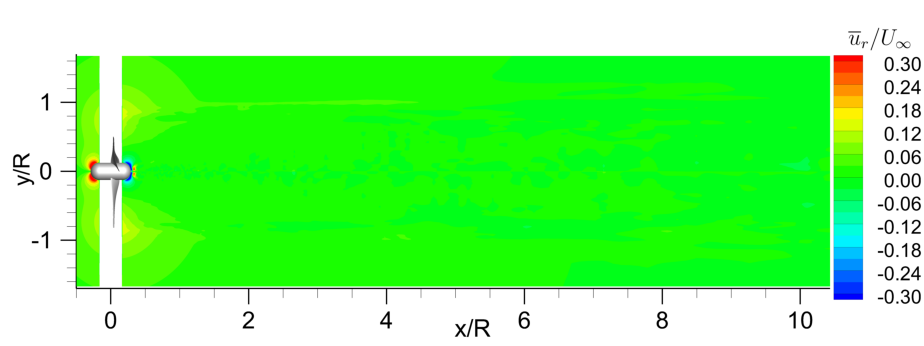}
		\caption{Open-rotor}
		\label{fig:owt_ave_u_radial}
	\end{subfigure}
	\caption{Time-averaged radial velocity contours in the central x-y plane}
	\label{fig:ave_u_radial}
\end{figure}

\begin{figure}[h!]
	\centering
	\begin{subfigure}[b]{0.48\textwidth}
		\centering
		\includegraphics[width=3.0in]{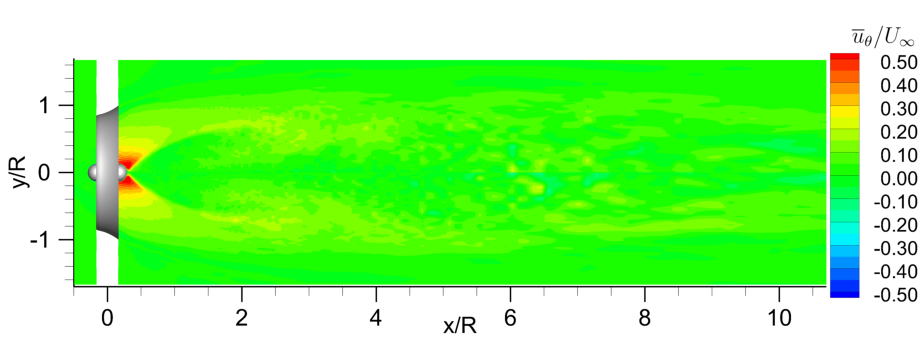}
		\caption{Dcuted}
		\label{fig:dwt_ave_u_theta}
	\end{subfigure}
	\begin{subfigure}[b]{0.48\textwidth}
		\centering
		\includegraphics[width=3.0in]{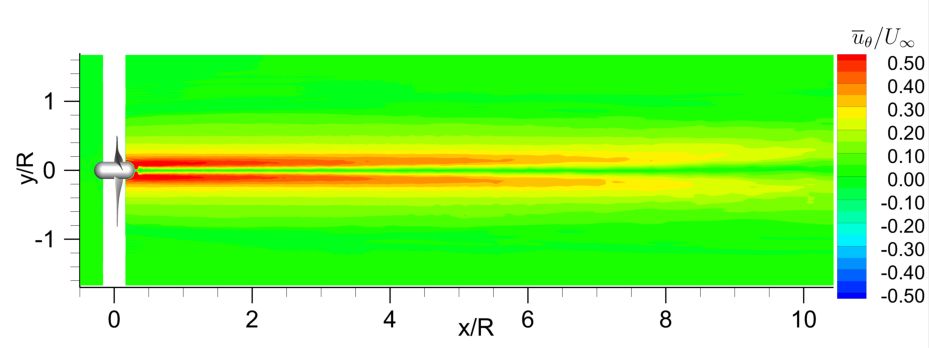}
		\caption{Open-rotor}
		\label{fig:owt_ave_u_theta}
	\end{subfigure}
	\caption{Time-averaged azimuthal velocity contours in the central x-y plane}
	\label{fig:ave_u_theta}
\end{figure}

The velocity vector can be decomposed into three components: streamwise velocity $u$, radial velocity $u_r$, and azimuthal velocity $u_{\theta}$.  Time-averaged values of these components are visualized in the central $x-y$ plane in Figs. \ref{fig:ave_u_stream}, \ref{fig:ave_u_radial}, and \ref{fig:ave_u_theta}.

As shown in Fig.\ref{fig:ave_u_stream}, both the ducted and open-rotor cases have an area with low streamwise velocity behind the hub end. For the ducted configuration, the low-velocity area expands in the radial direction as the flow moves downstream. After $x=4R$, this area begins to shrink. On the contrary, the low-speed area of the open-rotor case almost keeps its radial size until $x=9R$, and then gradually vanishes, of which the shape is  similar as the hub vortex shown in Fig. \ref{fig:owt_Q_ins}.

Fig. \ref{fig:dwt_ave_u_radial} presents the time-averaged radial velocity contour  for the ducted case. The area around the duct has a large positive radial velocity because the duct's diverging shape make the flow travels away from the central axis. Another high radial velocity region appears near the hub end, which means that the flow expands after it moves past the end. For the open-rotor case, there is no high radial velocity region like the ducted one.

Fig. \ref{fig:ave_u_theta} demonstrates the time-averaged azimuthal velocity fields in the x-y plane. In the open-rotor case, two high azimuthal velocity zones near the central line appear, which means that at this place the flow rotates quickly around the central axis, and is consistent with the hub vortex seen in Fig. \ref{fig:owt_Q_ins}. For the ducted case, only the flow around the hub end rotates quickly.

\begin{figure}[h!]
	\centering
	\begin{subfigure}[b]{0.19\textwidth}
		\centering
		\includegraphics[width=1.2in]{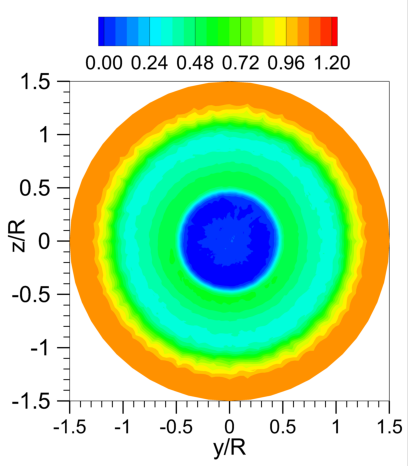}
		\caption{x=R}
		\label{fig:dwt_ave_u_stream_1}
	\end{subfigure}
	\begin{subfigure}[b]{0.19\textwidth}
		\centering
		\includegraphics[width=1.2in]{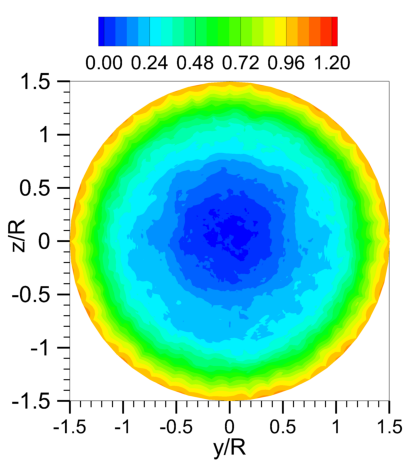}
		\caption{x=3R}
		\label{fig:dwt_ave_u_stream_3}
	\end{subfigure}
	\begin{subfigure}[b]{0.19\textwidth}
		\centering
		\includegraphics[width=1.2in]{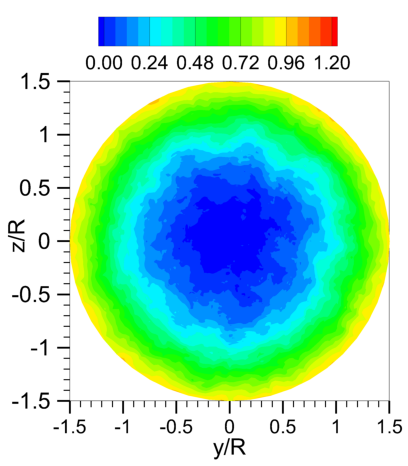}
		\caption{x=5R}
		\label{fig:dwt_ave_u_stream_5}
	\end{subfigure}
		\begin{subfigure}[b]{0.19\textwidth}
		\centering
		\includegraphics[width=1.2in]{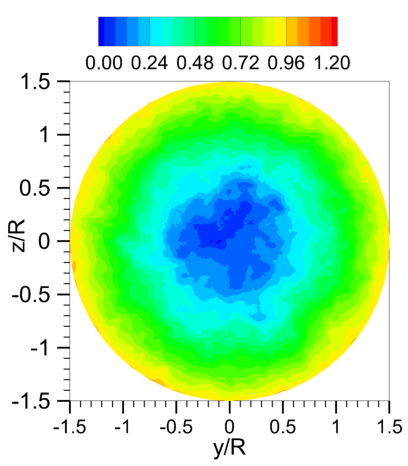}
		\caption{x=7R}
		\label{fig:dwt_ave_u_stream_7}
	\end{subfigure}
	\begin{subfigure}[b]{0.19\textwidth}
		\centering
		\includegraphics[width=1.2in]{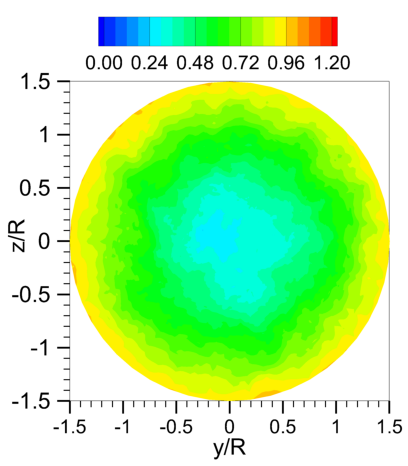}
		\caption{x=9R}
		\label{fig:dwt_ave_u_stream_9}
	\end{subfigure}
	\caption{Time-averaged streamwise velocity contours for the ducted case at different streamwise locations}
	\label{fig:dwt_ave_u_stream_locations}
\end{figure}
\begin{figure}[h!]
	\centering
	\begin{subfigure}[b]{0.19\textwidth}
		\centering
		\includegraphics[width=1.2in]{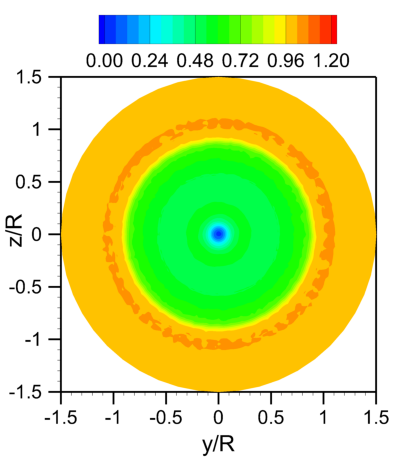}
		\caption{x=R}
		\label{fig:owt_ave_u_stream_1}
	\end{subfigure}
	\begin{subfigure}[b]{0.19\textwidth}
		\centering
		\includegraphics[width=1.2in]{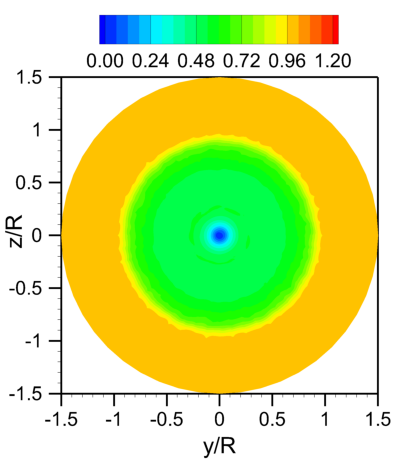}
		\caption{x=3R}
		\label{fig:owt_ave_u_stream_3}
	\end{subfigure}
	\begin{subfigure}[b]{0.19\textwidth}
		\centering
		\includegraphics[width=1.2in]{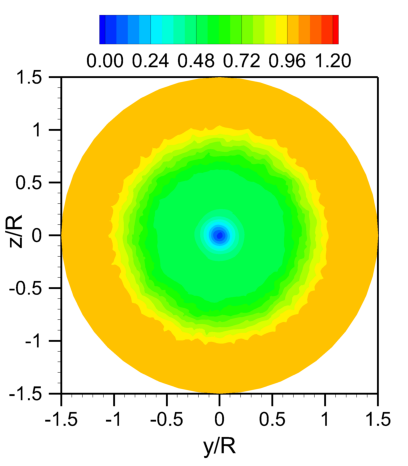}
		\caption{x=5R}
		\label{fig:owt_ave_u_stream_5}
	\end{subfigure}
	\begin{subfigure}[b]{0.19\textwidth}
		\centering
		\includegraphics[width=1.2in]{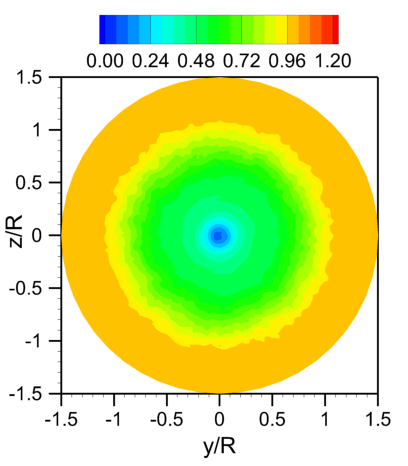}
		\caption{x=7R}
		\label{fig:owt_ave_u_stream_7}
	\end{subfigure}
	\begin{subfigure}[b]{0.19\textwidth}
		\centering
		\includegraphics[width=1.2in]{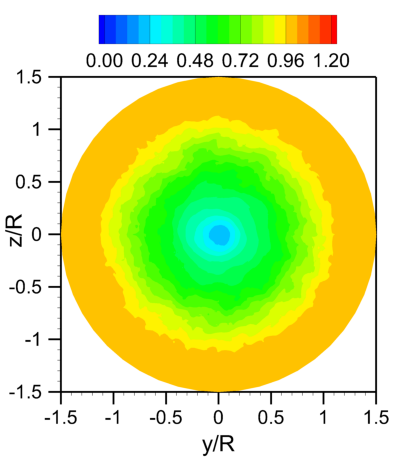}
		\caption{x=9R}
		\label{fig:owt_ave_u_stream_9}
	\end{subfigure}
	\caption{Time-averaged streamwise velocity contours for the open-rotor case at different streamwise locations.}
	\label{fig:owt_ave_u_stream_locations}
\end{figure}

To obtain more details of the streamwise velocity field, we plot the time-averaged  streamwise velocity contours at five different streamwise locations for the ducted and open-rotor configurations in Figs. \ref{fig:dwt_ave_u_stream_locations} and \ref{fig:owt_ave_u_stream_locations}, respectively. From Fig. \ref{fig:dwt_ave_u_stream_1}, it can be observed that a circular low-velocity region at the central part, which are circumscribed by two ring regions with higher velocity.
The central circular area corresponds to the wake just behind the hub end, and the inner ring region corresponds to the wake behind the area between the duct and the hub. As the flow moves downstream, boundaries between these regions become unclear due to the momentum exchange between different regions. At $x=9R$, these areas are totally mixed together.
In Fig. \ref{fig:owt_ave_u_stream_locations}, we can also see three regions for the open-rotor configuration. However, the area of the central low-velocity circular region is much less than that in Fig.\ref{fig:dwt_ave_u_stream_locations}, The inner ring region is formed by the flow past the rotor's blades. The central area keeps its shape until x=9R. The boundary between two ring areas become obscure after $x=3R$. However, these region are not blended together even when the flow moves downstream to $x=9R$.
One explanation for this difference is that the ducted turbine's wake is filled with small flow structures, and thus has a higher rate of momentum exchange, which leads to the difference of velocity between regions disappear more quickly.

\begin{figure}[h!]
	\centering
	\begin{subfigure}[b]{0.19\textwidth}
		\centering
		\includegraphics[width=1.2in]{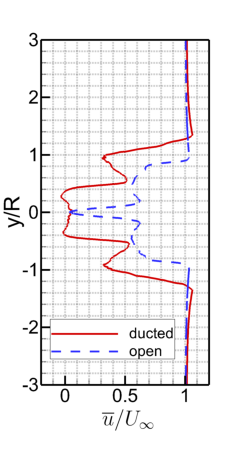}
		\caption{x=R}
		\label{fig:u_profile_1}
	\end{subfigure}
	\begin{subfigure}[b]{0.19\textwidth}
		\centering
		\includegraphics[width=1.2in]{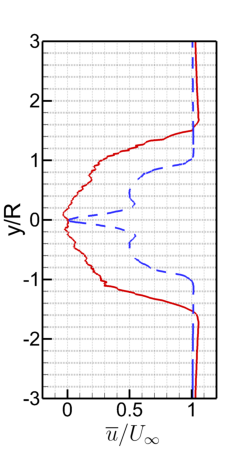}
		\caption{x=3R}
		\label{fig:u_profile_3}
	\end{subfigure}
	\begin{subfigure}[b]{0.19\textwidth}
		\centering
		\includegraphics[width=1.2in]{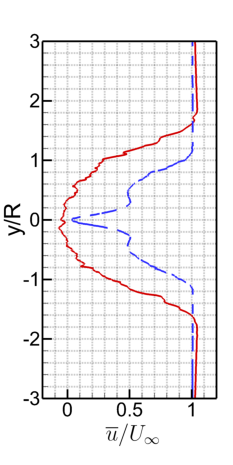}
		\caption{x=5R}
		\label{fig:u_profile_5}
	\end{subfigure}
	\begin{subfigure}[b]{0.19\textwidth}
		\centering
		\includegraphics[width=1.2in]{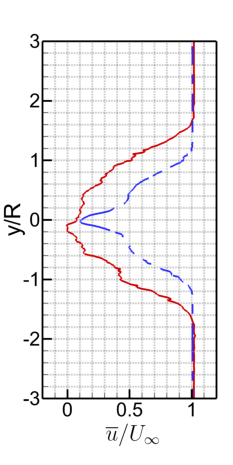}
		\caption{x=7R}
		\label{fig:u_profile_7}
	\end{subfigure}
	\begin{subfigure}[b]{0.19\textwidth}
		\centering
		\includegraphics[width=1.2in]{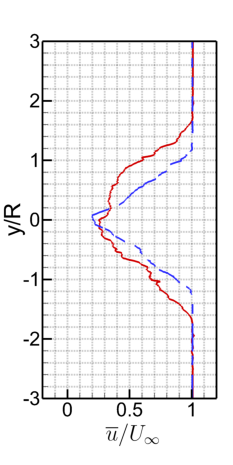}
		\caption{x=9R}
		\label{fig:u_profile_9}
	\end{subfigure}
	\caption{Time-averaged streamwise velocity profiles  at different streamwise locations in the central x-y plane.}
	\label{fig:ave_u_profile}
\end{figure}

To quantitatively compare the two configurations' wakes, time-averaged streamwise velocity profiles are plotted in Fig. \ref{fig:ave_u_profile}. Overall these profiles are symmetrical about $y=0$. The wake behind the ducted turbine has more velocity loss. When $x=R$,  the ducted configuration has an almost stationary flow in the area $-0.5\leq y/R \leq 0.5$, which corresponds to the hub wake. As $\abs{y}$ increases, the velocity of the ducted turbine increases quickly, and then has two local maximum $\overline{u}/U_{\infty}=0.3$ at $y/R\approx 1$, i.e. the duct exit radius.
The flow at $x=R$ continues to recover as $\abs{y}$ increase, and reach the freestream velocity $\overline{u}/U_{\infty}=1$ when $\abs{y}/R\geq 1.2$. By contrast, the open-rotor case has a spike around $x=0$ with a local minimum $\overline{u}/U_{\infty}=0$, and recovers quickly to  $\overline{u}/U_{\infty}=1$ when $\abs{y}/R\geq 1$.

When the flow moves downstream, the ducted case's velocity profiles has only one minimum for $x\geq 3R$, and is still smaller than that of the open-rotor one when $x\leq 7R$. When $x=9R$, the two profiles have a similar shape with a minimum value $\overline{u}/U_{\infty}\approx0.2$. Among other regions, the ducted case's velocity is smaller than the open-rotor one. The open-rotor configuration keeps its spike at $y=0$ until $x=7R$. The locations where two profiles reach the steady state remain unchanged.

\subsection{Yawed Flow}
\begin{figure}[h!]
	\centering
	\begin{subfigure}[b]{0.4\textwidth}
		\centering
		\includegraphics[width=2.in]{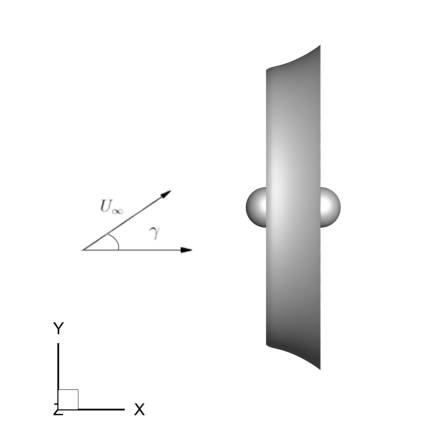}
		\caption{Yaw angle}
	\end{subfigure}
	\begin{subfigure}[b]{0.4\textwidth}
		\centering
		\includegraphics[width=2.0in]{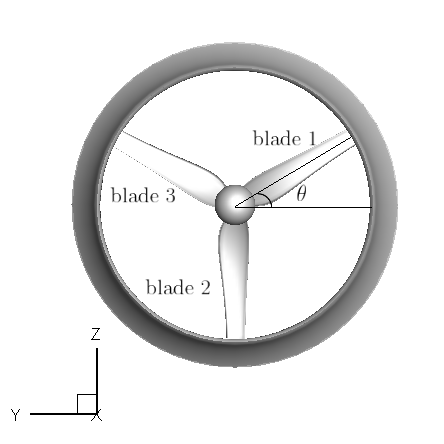}
		\caption{Azimuthal angle}
	\end{subfigure}
	\caption{Two angle definitions0 for yawed flow.}
	\label{fig:dwt_geo_yaw}
\end{figure}

\begin{figure}[h!]
	\centering
	\begin{subfigure}[b]{0.4\textwidth}
		\includegraphics[width=2.5in]{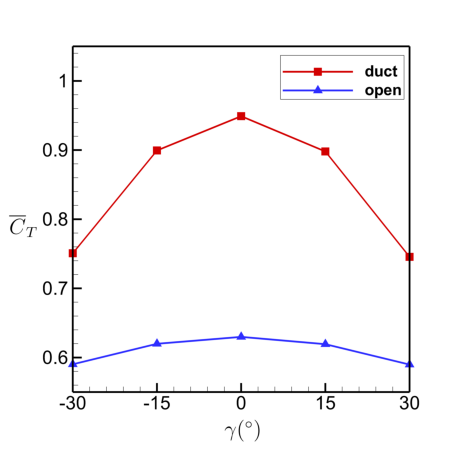}
		\caption{$C_T$}
		\label{fig:yaw_ct_non_norm}
	\end{subfigure}
	\begin{subfigure}[b]{0.4\textwidth}
		\includegraphics[width=2.5in]{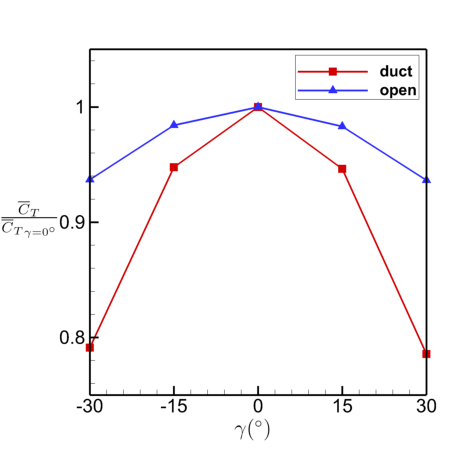}
		\caption{Normalized $C_T$}
		\label{fig:yaw_ct_norm}
	\end{subfigure}
	\caption{Time-averaged $C_T$ for two configurations under yawed flow.}
	\label{fig:yaw_ct}
\end{figure}

This section focuses on the yawed flows. To facilitate the description, two geometric parameters: yaw angle and azimuthal angle, are defined and shown in Fig. \ref{fig:dwt_geo_yaw}. In this study, yaw angle is denoted as the angle between the x-axis and the flow direction in the central x-y plane. Azimuthal angle is defined as angle between the blade axis and the horizontal line shown in Fig.\ref{fig:dwt_geo_yaw}. Initial azimuthal angles for three blades are $30^{\circ}$, $150^{\circ}$, and $270^{\circ}$.
For the yawed flow, we consider five different yaw angles: $0^{\circ}$, $\pm 15^{\circ}$, and $\pm 30^{\circ}$ for two configurations.  Fig \ref{fig:yaw_ct} compares the time-averaged $C_T$ of two configurations for different yawed conditions. It can be seen that both the ducted turbine and the open-rotor turbine's $\overline{C}_T$ are symmetrical about $\gamma=0^{\circ}$, and decrease with the growing yaw angle magnitude. Even though the ducted turbine experiences a larger drop on $\overline{C}_T$(about $0.2$) from $0^{\circ}$ to $30^{\circ}$ than the open-rotor one (about $0.04$), the ducted rotor experiences a larger thrust on every yawed angle. To compare the relative change, in Fig. \ref{fig:yaw_ct_norm} we normalize both configurations' $\overline{C}_T$ by their own $\overline{C}_T$ at $\gamma = 0^{\circ}$.  It can be found that relative changes of $\overline{C}_T$ for the open-rotor cases are smaller than that of the ducted ones. Especially, for $\gamma=\pm 30^{\circ}$, the open rotor only experiences about $6\%$ thrust drop, much less than the 20$\%$ for the ducted case.

\begin{figure}[H]
	\centering
	\begin{subfigure}[b]{0.4\textwidth}
		\includegraphics[width=2.5in]{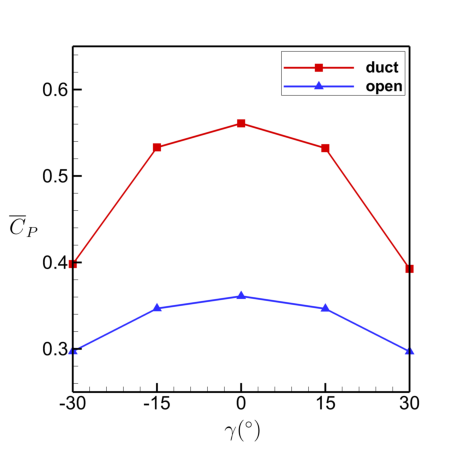}
		\caption{$C_P$}
		\label{fig:yaw_cp_non_norm}
	\end{subfigure}
	\begin{subfigure}[b]{0.4\textwidth}
		\includegraphics[width=2.5in]{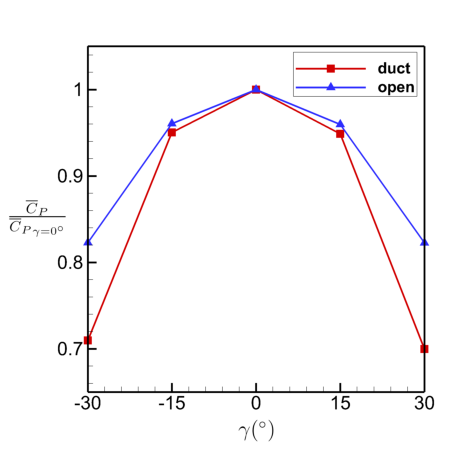}
		\caption{Normalized $C_P$}
		\label{fig:yaw_cp_norm}
	\end{subfigure}
	\caption{Time-averaged $C_P$ for the two configurations under yawed flows.}
	\label{fig:yaw_cp}
\end{figure}

\begin{figure}[h!]
	\centering
	\begin{subfigure}[b]{0.4\textwidth}
		\includegraphics[width=2.5in]{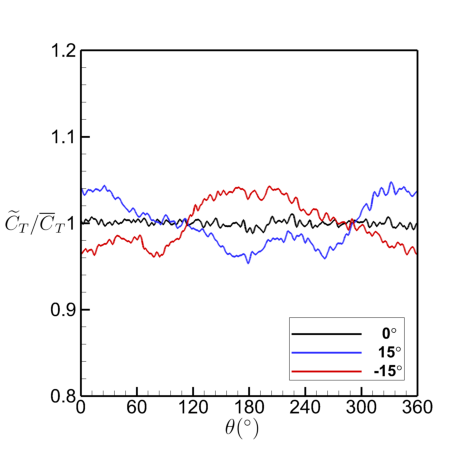}
		\caption{Ducted}
		\label{fig:dwt_ct_ave_1}
	\end{subfigure}
	\begin{subfigure}[b]{0.4\textwidth}
		\includegraphics[width=2.5in]{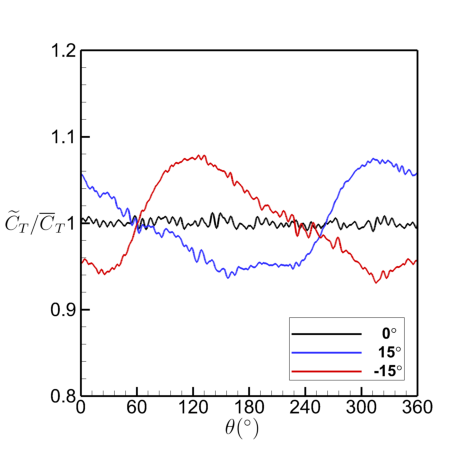}
		\caption{Open-rotor}
		\label{fig:dwt_ct_ave_2}
	\end{subfigure}
	\caption{Azimuthal-averaged $C_T$ of the blade 1 for yaw angle $\gamma=0,\pm15^{\circ}$.}
	\label{fig:yaw_ct_azimuthal_1}
\end{figure}

\begin{figure}[h!]
	\centering
	\begin{subfigure}[b]{0.4\textwidth}
		\includegraphics[width=2.5in]{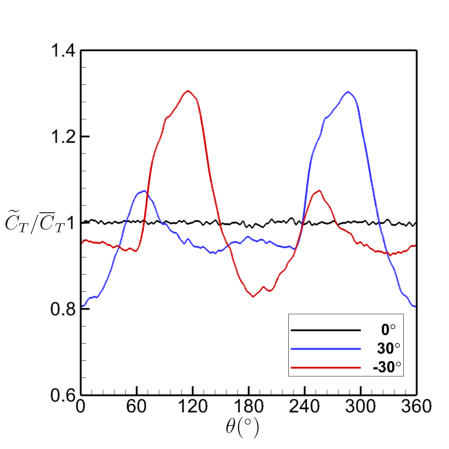}
		\caption{Ducted}
		\label{fig:owt_ct_ave_1}
	\end{subfigure}
	\begin{subfigure}[b]{0.4\textwidth}
		\includegraphics[width=2.5in]{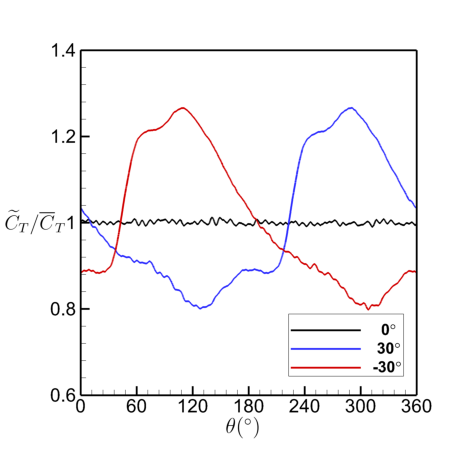}
		\caption{Open-rotor}
		\label{fig:owt_ct_ave_2}
	\end{subfigure}
	\caption{Azimuthal-averaged $C_T$ of the blade 1  for yaw angle $\gamma=0,\pm 30^{\circ}$.}
	\label{fig:yaw_ct_azimuthal_2}
\end{figure}

As shown in Fig. \ref{fig:yaw_cp}, the $\overline{C}_P$ curves for yawed conditions have a similar trend as the $\overline{C}_T$ curves, which have a symmetry about $\gamma=0^{\circ}$ and see a drop as the yaw angle grows. The ducted turbine has more power output than the open rotor in all yawed cases, and experiences a drop by $0.12$ on its $\overline{C}_P$ value when $\gamma=\pm30^{\circ}$. In comparison, the open rotor only decreases by $0.05$.  In terms of the relative change in $\overline{C}_P$, the ducted configuration has a much larger drop (about $30\%$) for $\gamma=\pm 30^{\circ}$, whereas the open-rotor case only drops by about $18\%$. The $\overline{C}_P$ values are symmetric with respect to $\gamma=0^{\circ}$, and decrease as $\gamma$ increases. The ducted rotor have more power output than the open rotor on all yawed angles. For $\gamma=\pm15^{\circ}$, relative changes for two configurations are similar (around 5\%). However, the ducted configuration has a much larger relative decrease on power output (about $30\%$) for $\gamma=\pm 30^{\circ}$. The open-rotor case only experiences around $18\%$ decrease on $\overline{C}_P$.
$\widetilde{C}_T$, i.e., the azimuthal-averaged $C_T$ can be computed by averaging $C_T$ values at the time instant when the azimuthal angle is equal to a given value, like $90^{\circ}$. Fig. \ref{fig:yaw_ct_azimuthal_1} shows $\widetilde{C}_T$ of blade 1 for $\gamma=0,\pm 15^{\circ}$, which is normalized by the corresponding $\overline{C}_T$. We see that all yawed cases have a sine-like fluctuation during one period. One major cause is the blade advancing and retreating effects reported in \cite{schulz-2017}. When the turbine blade rotates in yawed flows, it will move upstream during half of the rotation period, and move downstream during the other half of the period, which results in the variation of the relative velocity and then the variation of angle of attack. Thus a fluctuation in thrust is seen during one revolution. When $\gamma = 0^{\circ}$, $C_T$ remain stable around $\overline{C}_T$ for both configurations. Both cases have a $180^{\circ}$ phase difference for $C_T$ values between $\gamma=15^{\circ}$ and $\gamma=-15^{\circ}$. For $\gamma = -15^{\circ}$, ducted cases has a maximum at $\theta=180^{\circ}$, whereas open-rotor cases  has its maximum at around $\theta=120^{\circ}$ and has a larger amplitude (about $1.1$).
As shown in Fig. \ref{fig:yaw_ct_azimuthal_2}, For $\gamma=\pm 30^{\circ}$, these $C_T$ curves have the same phase difference between the positive yaw angle and the negative yaw angle.  Both configurations' maximum values appear at $\theta=120^{\circ}$. However the ducted one has a larger amplitude (about $1.3$), and has another local maximum at around $\theta=250^{\circ}$ due to the duct's effects.

\section{Conclusion}
In this study, we perform high-order large eddy simulations for a wind turbine under two configurations: ducted and open-rotor.

For the axial flow situation, a comparison of the thrust coefficient $C_T$ and the power coefficient $C_P$ between the two cases is made. It is found that the ducted one has higher mean values for thrust and power output than its bare counterpart, and also has higher r.m.s deviation values, which means stronger load fluctuations. To better understand the load distribution, time-averaged pressure fields on the blade surfaces and near-blade regions. We see that the two configurations experience a drag force pointed towards the x axis, and a torque pointed towards the negative x axis, and the load concentrates on the outboard part of the blades, especially around the leading edges. To investigate the wake dynamics, Q-criterion isosurfaces are plotted. It is found that the wake of the ducted turbine is filled with small flow structures, whereas the open rotor has a coherent hub vortex and weak tip vortices that are dissipated quickly by the flow. The time-averaged velocity contours for three components: $u$, $u_r$, and $u_{\theta}$  are also presented. We found that the ducted case has a larger low-velocity wake area behind the hub end. In comparison, the open-rotor configuration has a thinner  band region with low streamwise velocity and high azimuthal velocity. Streamwise velocity contours at different streamwise locations reveal that the wake behind the ducted rotor has a faster rate of momentum exchange.

For the yawed flow situations, we compare $C_T$ and $C_P$ of the two configurations. For all tested yaw angles, the ducted one has a higher thrust and power output than its bare counterpart. However, the ducted case experiences a larger drop on $C_T$ and $C_P$ in both absolute and relative senses. We also compute $\widetilde{C}_T$, i.e.,  the azimuthal-averaged $C_T$, for one blade. Due to the blade advancing and retreating effect, each yawed case's curve has some fluctuations. And there is a $180^{\circ}$ phase difference between two flows with the same magnitude but opposite signs.

To better understand the performance of the ducted wind turbine, we will perform simulations for different tip speed ratios in the future.

\section*{Acknowledgments}
The authors would like to express our acknowledgments to Clarkson University for financial support through the IGNITE Fellowship Program.
\bibliography{automata}

\end{document}